\pdfoutput=1
\documentclass[12pt,twoside,english,3p]{elsarticle}
\usepackage{fourier}
\usepackage[T1]{fontenc}
\usepackage[latin9]{inputenc}
\usepackage{color}
\usepackage{babel}
\usepackage{booktabs}
\usepackage{bm}
\usepackage{amsmath}
\usepackage{amssymb}
\usepackage{graphicx}
\usepackage{esint}
\usepackage[unicode=true,
 bookmarks=true,bookmarksnumbered=true,bookmarksopen=false,
 breaklinks=false,pdfborder={0 0 0},pdfborderstyle={},backref=false,colorlinks=true]
 {hyperref}
\hypersetup{pdftitle={A minimal hyperbolic system for unstable shock waves},
 pdfauthor={Dmitry I. Kabanov and Aslan R. Kasimov},
 pdfkeywords={hyperbolic systems, shock waves, stability, bifurcations, chaos, detonation},
 urlcolor=blue,linkcolor=blue,citecolor=blue}

\makeatletter

\providecommand{\tabularnewline}{\\}

\usepackage{babel}
\journal{Communications in Nonlinear Science and Numerical Simulation}

\usepackage{siunitx}

\usepackage[all]{hypcap}

\makeatother

\begin{document}
\begin{frontmatter}

\title{A minimal hyperbolic system for unstable shock waves}

\author[kaust]{Dmitry I.\ Kabanov}

\ead{dmitry.kabanov@kaust.edu.sa}

\author[lpi,sechenov]{Aslan R.\ Kasimov\corref{cor}}

\ead{kasimov@lpi.ru}

\cortext[cor]{Corresponding author}

\address[kaust]{Division of Computer, Electrical and Mathematical Sciences and Engineering
\\
King Abdullah University of Science and Technology, Box 4700, Thuwal
23955-6900, Saudi Arabia}

\address[lpi]{Tamm Theory Department, Lebedev Physical Institute \\
Russian Academy of Sciences, Leninsky prospekt, 53, Moscow, 119991,
Russia}

\address[sechenov]{Sechenov University, Trubetskaya street, 8, stroenie 2, Moscow 119991,
Russia}
\begin{abstract}
We present a computational analysis of a 2$\times$2 hyperbolic system
of balance laws whose solutions exhibit complex nonlinear behavior.
Traveling-wave solutions of the system are shown to undergo a series
of bifurcations as a parameter in the model is varied. Linear and
nonlinear stability properties of the traveling waves are computed
numerically using accurate shock-fitting methods. The model may be
considered as a minimal hyperbolic system with chaotic solutions and
can also serve as a stringent numerical test problem for systems of
hyperbolic balance laws.
\end{abstract}
\end{frontmatter}

\global\long\def\R{\mathbb{R}}

\global\long\def\DeltaX{\Delta x}

\global\long\def\DeltaT{\Delta t}

\global\long\def\T{\mathrm{T}}

\global\long\def\shk{\mathrm{s}}

\global\long\def\Nhrz{N_{1/2}}

\global\long\def\TFinal{T_{\text{final}}}

\global\long\def\tolLambda{\text{tol}_{\lambda}}

\global\long\def\DCJ{D_{\text{CJ}}}

\global\long\def\bigO{\mathcal{O}}

\global\long\def\diag{\operatorname{diag}}

\global\long\def\vec#1{\bm{#1}}

\global\long\def\mat#1{\bm{#1}}

\section{Introduction\label{sec:fsm:intro}}

We investigate a particular hyperbolic system of balance laws in one
space dimension: 
\begin{eqnarray}
u_{t}+\left(f\left(u,\lambda\right)\right)_{x} & = & 0,\label{eq:balance-law}\\
\lambda_{t} & = & \omega\left(u,\lambda\right),\quad t>0,x\in\mathbb{R},\label{eq:rate-law}
\end{eqnarray}
where $u$ and $\lambda$ are the unknown functions, $f$ and $\omega$
are given flux and rate function, respectively, and subscripts $t$
and $x$ denote partial derivatives in time and space, respectively.
We demonstrate numerically that the system possesses nontrivial dynamical
properties. In particular, we show that its traveling-wave solutions
can become unstable as a system parameter is varied, and that the
instability manifests itself as an Andronov\textendash Hopf bifurcation
leading to a limit-cycle attractor. Further increase of the parameter
results in a cascade of period-doubling bifurcations and the onset
of apparently chaotic dynamics. 

Much analysis of systems of the type (\ref{eq:balance-law}-\ref{eq:rate-law})
is due to \citet{Fickett1979,fickett1984shock,fickett1985stability,Fickett1985},
who was the first to introduce it as an analog (or a toy model) of
the reactive Euler equations of gas dynamics with the purpose of modeling
the dynamics of detonations. A similar model which also included diffusive
effects was proposed independently by Majda \citep{Majda1980}. The
Majda model has received much attention in the mathematics literature
\citep{zumbrun2017recent,lyng2004stability,humpherys2013stability,humpherys2010efficient,levy1992majda}
as a prototype to study existence and stability of traveling waves.
It must be pointed out that in the analyses of the Majda model only
stable traveling waves have been found so far, to the best of our
knowledge. In contrast, the model studied in the present work predicts
instabilities, as pointed out originally in \citep{FariaKasimovRosales-JFM2015}.

The Fickett model \citep{Fickett1979} has subsequently motivated
various extensions and modifications \citep{RadulescuPRL2011,tang2012dynamics,kasimovPRL2013,FariaKasimovRosales-SIAM2014,FariaKasimovRosales-SIAM2016,FariaKasimov-PROCI2014}.
The principal aim of all of these works is to identify a minimal model
that is capable of reproducing the rich set of dynamical properties
of the full system of the reactive Euler equations. It is hoped that
doing so might help in revealing the key mechanisms of the observed
complex dynamics of the full system. As the recent publications mentioned
above have demonstrated, the Fickett model indeed successfully reproduces
most of the features of the full system. These results have also motivated
the development of an asymptotic theory of gaseous detonations in
\citep{FariaKasimovRosales-JFM2015} in which a reduced model is derived
that is found to be very similar to the Fickett \emph{ad hoc} model
(\ref{eq:balance-law}-\ref{eq:rate-law}), however with a difference
in the second equation \textendash{} instead of $\lambda_{t}$, the
asymptotic model has $\lambda_{x}$ (see also the earlier related
work \citep{rosales1989diffraction,RosalesMajda1983,clavin2002dynamics}).
It was stated in \citep{FariaKasimovRosales-JFM2015} that in either
case, the system possesses instabilities as long as the rate function
$\omega$ is chosen that has the right properties. Further analysis
of the system with $\lambda_{t}$ was not pursued by the authors of
\citep{FariaKasimovRosales-JFM2015}. Here, we carry out a complete
numerical investigation of such a system using the particular rate
function $\omega$ from the asymptotic model of \citep{FariaKasimovRosales-JFM2015}
as an example.

We also propose that (\ref{eq:balance-law}-\ref{eq:rate-law}) can
serve as a numerical benchmark problem for systems of hyperbolic balance
laws. Despite its simplicity, the system exhibits rather complex and
sensitive dynamics of solutions. As such, it can be used as a stringent
test problem for numerical algorithms that are to accurately reproduce
stability properties in problems with complex dynamical features.
As examples of such problems we mention detonations, shallow water
flows over topographies, shock waves in the presence of body forces
(e.g., gravitational or electromagnetic fields). A good numerical
method must correctly reproduce neutral stability boundaries and development
of instability as the boundary is crossed. For such problems, our
model and the results reported here can be used as a relatively simple
benchmark case.

The remainder of the paper is structured as follows. The model system
and its main mathematical properties are introduced in Section \ref{sec:model}.
The traveling-wave solutions of the model are found in Section \ref{sec:znd}.
The numerical algorithms used to calculate both linear and nonlinear
dynamics of the system require the so-called shock-evolution equation,
which is derived in Section \ref{sec:sheveq}. The linear stability
of traveling waves and nonlinear dynamics are presented in Sections
\ref{sec:lin} and \ref{sec:nonlinear}, respectively, while code
verification results are given in Section \ref{sec:verif}. Conclusions
are presented in Section \ref{sec:conclusions}.

\section{The model system\label{sec:model}}

In the original paper \citep{Fickett1979}, Fickett proposed a simple
\emph{ad hoc} system of hyperbolic equations to \emph{qualitatively}
model the dynamics of detonation waves. To remind the reader, a detonation
is a self-sustained shock wave propagating in a reactive medium such
that the shock compression and heating triggers exothermic chemical
reactions, and the thermal energy released in these reactions serves
to support the motion of the shock \citep{FickettDavis2011}. Usually,
detonations are modeled by the reactive Euler equations of gas dynamics
which consist of conservation laws of mass, momentum and energy, and
at least one equation that describes chemical heat release. Thus,
in one spatial dimension, this is a hyperbolic system of at least
four quasilinear equations. Analysis of detonations by means of the
reactive Euler equations has received much attention and continues
actively at present. See, for example, recent reviews \citep{bdzil2007dynamics,bdzil2012theory,stewart2006state,clavin2012analytical,zumbrun2017recent,Clavin-CST-2017}.
One of the key properties of gaseous detonations is their manifestly
time-dependent and spatially complex dynamics. Thus arises the problem
of understanding the physical mechanisms of such behavior and of the
mathematical properties of the governing equations that are responsible
for the observed complex dynamics.

Fickett's model system can be written as follows:
\begin{align}
u_{t}+\frac{1}{2}\left(u^{2}+q\lambda\right)_{x} & =0,\label{eq:model-lab-Burgers}\\
\lambda_{t} & =\omega\left(u,\lambda\right).\label{eq:model-lab-reaction}
\end{align}
Here, the first equation is to play the role of the combined momentum\textendash energy
equation while the second is the rate equation for the chemical energy
release, $q$ is the heat release parameter. The first equation of
the model is a conservation law with a flux function $f\left(u,\lambda\right)$
in which the second variable $\lambda$ satisfies the ordinary differential
equation (ODE) \eqref{eq:model-lab-reaction}. The variable $\lambda$
measures the reaction progress, varying from $\lambda=0$ in the unburnt
(ambient) state ahead of the shock to $\lambda=1$ in the burnt state
far downstream of the shock. 

We assume that the shock moves from left to right in the positive
$x$ direction. Thus, if $x=x_{\shk}\left(t\right)$ denotes the shock
position at time $t$, the upstream unburnt state is at $x>x_{\shk}\left(t\right)$,
the reaction zone is at $x<x_{\shk}\left(t\right)$, and the burnt
state is reached asymptotically at $x\to-\infty$.

We choose the reaction rate following \citep{FariaKasimovRosales-JFM2015}
as 
\begin{equation}
\omega(u,\lambda)=\begin{cases}
k(1-\lambda)\exp\left(\theta\left(\sqrt{q}u+q\lambda\right)\right), & x<x_{\shk}\left(t\right),\\
0, & x>x_{\shk}\left(t\right),
\end{cases}\label{eq:model:reaction-rate}
\end{equation}
where $k$ is the rate constant, $\theta$ is the activation energy,
and $q$ is the same heat release parameter as in \eqref{eq:model-lab-Burgers}.
This form of the rate function was derived from the reactive Euler
equations in a particular asymptotic limit of weakly nonlinear waves
in \citep{FariaKasimovRosales-JFM2015}. It must be emphasized however
that our use of the function is outside the asymptotic theory and
must be considered only as a particular case of Fickett's analog system. 

In vector form, system (\ref{eq:model-lab-Burgers}\textendash \ref{eq:model-lab-reaction})
can be written as 
\begin{equation}
\vec z_{t}+\left(\vec f\left(\vec z\right)\right)_{x}=\vec s\left(\vec z\right),\label{eq:model-system}
\end{equation}
where 
\begin{align*}
\vec z=\left[\begin{array}{l}
u\\
\lambda
\end{array}\right],\quad\vec f=\left[\begin{array}{l}
\frac{1}{2}u^{2}+\sigma\lambda\\
0
\end{array}\right],\quad\vec s=\left[\begin{array}{l}
0\\
\omega
\end{array}\right],
\end{align*}
and $\sigma=q/2$ will also be used along with $q$. The Jacobian
of $\vec f$, 
\begin{equation}
\mathbf{A}=\left[\begin{array}{ll}
u & \sigma\\
0 & 0
\end{array}\right],
\end{equation}
has eigenvalues $\mu_{1}=u$ and $\mu_{2}=0$ with corresponding right
eigenvectors $\vec r_{1}=[1,0]^{\T}$ and $\vec r_{2}=[1,-u/\sigma]^{\T}$.
Clearly, the second characteristic field is linearly degenerate while
the first is genuinely nonlinear as $\nabla\mu_{1}\cdot\vec r_{1}=1\neq0$.

In nonconservative form, the system (\ref{eq:model-lab-Burgers}\textendash \ref{eq:model-lab-reaction})
becomes
\begin{align}
u_{t}+uu_{x}+\sigma\lambda_{x} & =0,\label{eq:model-lab-noncon-Burgers}\\
\lambda_{t} & =\omega,\label{eq:model-lab-noncon-reaction}
\end{align}
from which the characteristic form of the system readily follows by
adding \eqref{eq:model-lab-noncon-reaction} multiplied by $\sigma$
to \eqref{eq:model-lab-noncon-Burgers} multiplied by $u$ (\eqref{eq:model-lab-noncon-reaction}
is already in the characteristic form):
\begin{alignat}{2}
\dot{p} & =\sigma\omega & \quad\text{on }\dot{x} & =u,\label{eq:char-u}\\
\dot{\lambda} & =\omega & \quad\text{on }\dot{x} & =0,\label{eq:char-lambda}
\end{alignat}
where $p=u^{2}/2+\sigma\lambda$, and the dots are used here and from
now on to denote total time derivatives, e.\ g., $\dot{p}=dp/dt$.

For the numerical solution of (\ref{eq:model-lab-Burgers}\textendash \ref{eq:model-lab-reaction}),
it is advantageous to move to a reference frame attached to the lead
shock by using new coordinates $\xi=x-x_{\shk}\left(t\right)$ and
$\tau=t$. In this new reference frame, the shock is always at the
same position, $\xi=0$, which serves as the right-end boundary of
the reaction zone. Because no finite-differencing is made across the
shock, the usual numerical shock smearing is removed in this formulation,
which is its key advantage. Reusing the notation $x$ and $t$ in
place of $\xi$ and $\tau$ and denoting the shock velocity as $D=\dot{x}_{\shk}$,
the governing equations in the shock-attached frame become: 
\begin{align}
u_{t}+\frac{1}{2}\left(\left(u-2D\right)u+q\lambda\right)_{x} & =0,\label{eq:model-Burgers}\\
\lambda_{t}-D\lambda_{x} & =\omega.\label{eq:model-reaction}
\end{align}

Equations (\ref{eq:model-Burgers}\textendash \ref{eq:model-reaction})
must be supplemented with the Rankine\textendash Hugoniot conditions
for $u$ and $\lambda$ at the shock (see e.g., \citep{leveque1992numerical}).
We denote the state variables right ahead of the shock (at position
$x_{\mathrm{s}}^{+}$) by subscript ``a'' (for ``ambient''), and
the state variables right behind the shock (at position $x_{\mathrm{s}}^{-}$)
by subscript ``s'' (for ``shock''). We also assume for simplicity
that the ambient (upstream) conditions ahead of the shock are $u_{\mathrm{a}}=0$
and $\lambda_{\mathrm{a}}=0$. The Rankine\textendash Hugoniot condition
for $\lambda$ is $\left[\lambda\right]=0$, where brackets denote
the jump across the shock ($\left[z\right]=z^{+}-z^{-}$), because
no reaction is assumed to take place inside the shock. As a result,
we obtain
\begin{equation}
\lambda_{\shk}=0.\label{eq:lambda_s}
\end{equation}
The jump in $u$ in \eqref{eq:model-Burgers} satisfies 
\begin{equation}
\left[\left(u-2D\right)u\right]+q\left[\lambda\right]=0,\label{eq:RH_u}
\end{equation}
which yields
\begin{equation}
u_{\mathrm{s}}=2D.\label{eq:u_s}
\end{equation}

\section{The traveling shock-wave solution\label{sec:znd}}

Next, we calculate the traveling-wave solutions of (\ref{eq:model-Burgers}\textendash \ref{eq:model-reaction})
(called ZND solutions after \citet{Zeldovich1940,vonNeumann1942,Doering1943}).
Substituting $u=\bar{u}\left(x-\bar{D}t\right)$, $\lambda=\bar{\lambda}\left(x-\bar{D}t\right)$
into the system, with overbar denoting the steady state and $\bar{D}=\text{const}$,
we obtain
\begin{align}
\frac{d}{dx}\left(\frac{\bar{u}^{2}}{2}-\bar{D}\bar{u}+\frac{q}{2}\bar{\lambda}\right) & =0,\label{eq:znd:eqn-of-motion}\\
\frac{d\bar{\lambda}}{dx} & =-\frac{\bar{\omega}}{\bar{D}}.\label{eq:znd:eqn-of-reaction}
\end{align}
From \eqref{eq:znd:eqn-of-motion}, we find the algebraic relation
\[
\bar{u}^{2}-2\bar{D}\bar{u}+q\bar{\lambda}=\text{const}=\bar{u}_{\shk}^{2}-2\bar{D}\bar{u}_{\shk}+q\bar{\lambda}_{\shk}=0\quad\text{(with }\bar{u}_{\shk}=2\bar{D}\text{)}.
\]
The root of this equation satisfying condition~\eqref{eq:u_s} is
\begin{equation}
\bar{u}=\bar{D}+\sqrt{\bar{D}^{2}-q\bar{\lambda}}.\label{eq:znd:u-of-lambda}
\end{equation}
Then, the spatial structure of the traveling-wave solution is found
by substituting \eqref{eq:znd:u-of-lambda} into \eqref{eq:znd:eqn-of-reaction}
and integrating the resultant ODE 
\begin{equation}
\frac{d\bar{\lambda}}{dx}=-\frac{\bar{\omega}\left(\bar{u}\left(\bar{\lambda}\right),\bar{\lambda}\right)}{\bar{D}},\label{eq:znd:dlambda_dx}
\end{equation}
from $x=0$ to $x<0$ using $\bar{\lambda}\left(0\right)=0$.

Importantly, so far $\bar{D}$ remains unknown. As in the classical
detonation theory \citep{FickettDavis2011}, the steady detonation
velocity is found using a special condition at a sonic point, where
the flow speed relative to the shock becomes equal to the local sound
speed. Equivalently, this sonic condition follows from the requirement
that the solution remains regular everywhere in the post-shock region.
For our particular case, this means that $\bar{u}\left(x\right)$
must remain sufficiently smooth everywhere at $x<0$. From \eqref{eq:znd:u-of-lambda}
and \eqref{eq:znd:dlambda_dx}, we find that 

\begin{align}
\frac{d\bar{u}}{dx} & =\frac{1}{2}\frac{-q}{\sqrt{\bar{D}^{2}-q\bar{\lambda}}}\frac{d\bar{\lambda}}{dx}=\frac{1}{2}\frac{q\bar{\omega}}{\bar{D}\sqrt{\bar{D}^{2}-q\bar{\lambda}}}.\label{eq:fsm:znd:du_dx}
\end{align}
Therefore, this derivative can blow up if $\bar{D}^{2}-q\bar{\lambda}=0$
at any point where $\bar{\omega}\neq0$. With our choice of the reaction
rate, $\bar{\omega}$ vanishes only at $\bar{\lambda}=1$. Hence,
to avoid any singularity in $\bar{u}$, we require that $\bar{D}^{2}-q\bar{\lambda}=0$
whenever $\bar{\lambda}=1$. This condition yields a unique value
for the traveling-wave speed (called the Chapman\textendash Jouguet,
or CJ speed in detonation theory \citep{FickettDavis2011}), 
\begin{equation}
\bar{D}=\sqrt{q}.\label{eq:D_CJ}
\end{equation}

The pre-exponential factor, $k$, in \eqref{eq:model:reaction-rate}
is a property of chemical reactions, related to the frequency of molecular
collisions. It is customary in detonation theory to eliminate this
parameter by rescaling the spatial coordinate such that a characteristic
length scale is the so-called half-reaction length of the steady reaction
zone, that is, the distance between the shock and the point where
half of the chemical energy is released, i.\,e., $\bar{\lambda}=1/2$.
Then, we find from \eqref{eq:znd:dlambda_dx} that with 
\begin{equation}
k=\int_{0}^{1/2}\frac{\bar{D}}{\left(1-\bar{\lambda}\right)\exp\left(\theta\left(\sqrt{q}\bar{u}+q\bar{\lambda}\right)\right)}\,d\bar{\lambda},\label{eq:znd:k}
\end{equation}
$x=-1$ where $\bar{\lambda}=1/2.$

\begin{figure}
\begin{centering}
\includegraphics{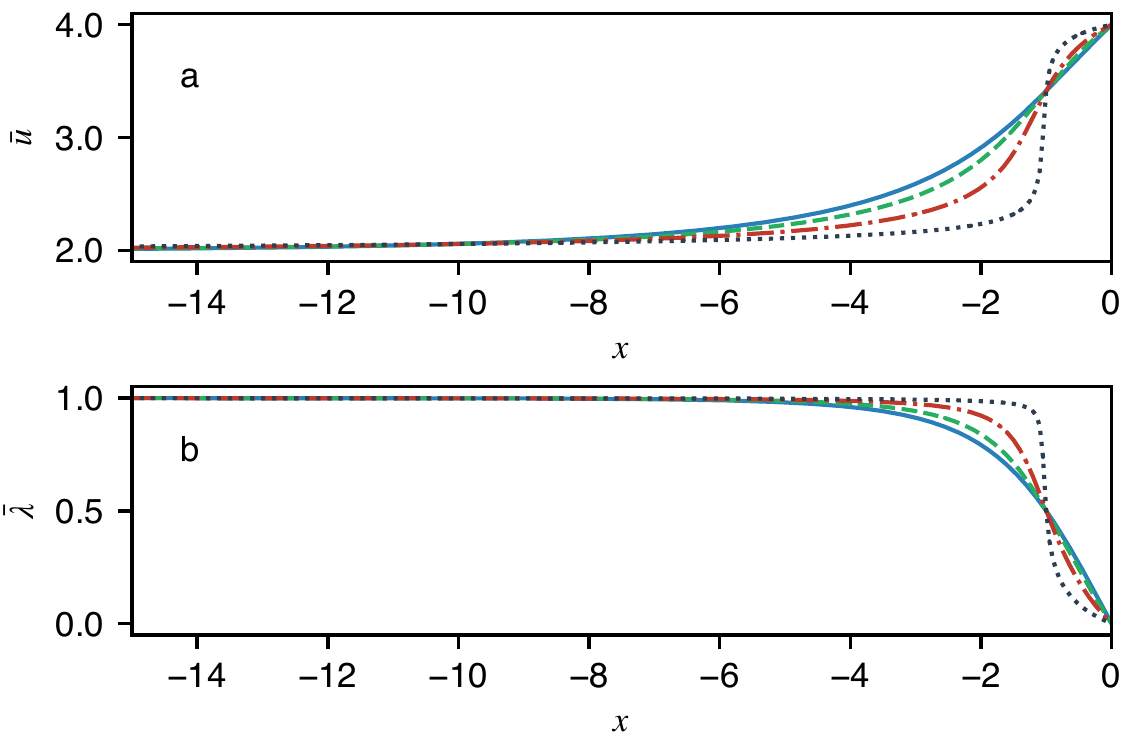}
\par\end{centering}
\caption{ZND profiles for $q=4$: a) velocity $\bar{u}$, b) reaction progress
variable $\bar{\lambda}$, as $\theta$ is varied: $\theta=0.5$ (solid
line), $\theta=1$ (dashed line), $\theta=2$ (dashed-dotted line),
$\theta=5$ (dotted line). With increase of $\theta$ ZND profiles
become steeper and approach ``square-wave'' structure, in which
reaction effectively occurs in a narrow region near the lead shock
at $x=0$.\label{fig:znd-solutions}}
\end{figure}

Figure \ref{fig:znd-solutions} shows computed spatial profiles of
$\bar{u}$ and $\bar{\lambda}$ at $q=4$ and varying activation energy
$\theta\in\left\{ 0.5,1,2,5\right\} $. The plot demonstrates that
as the value of the activation energy increases, the profile of $\bar{\lambda}$
develops a steeper slope near $x=-1$, implying an increased stiffness
of the problem at large $\theta$. 

\section{Shock-evolution equation\label{sec:sheveq}}

The detonation speed, $D$, which is generally unknown for time-dependent
solutions, appears explicitly in (\ref{eq:model-Burgers}-\ref{eq:model-reaction})
in the shock-attached frame. Therefore, a method is needed to compute
$D$ when solving the system numerically. As in the related previous
work \citep{HenrickAslamPowers2006,kasimov2004dynamics,taylor2009mode},
in this subsection, we derive a shock-evolution equation (also called
the ``shock-change equation'' \citep{ChenGurtin1971,Fickett1985})
that is used subsequently to determine $D$ as part of the numerical
algorithms that follow.

The evolution equation is derived by computing the rate of change
of $u$ on the shock path by two different means. On the one hand,
using the laboratory-frame equations (\ref{eq:model-lab-Burgers}-\ref{eq:model-lab-reaction}),
we find that
\begin{equation}
\dot{u}|_{\mathrm{s}}=\left(u_{t}+Du_{x}\right)|_{\mathrm{s}},
\end{equation}
and eliminating $u_{t}$ from here using \eqref{eq:model-lab-Burgers},
it follows that 
\begin{equation}
\dot{u}|_{\mathrm{s}}=\left((D-u)u_{x}-\sigma\lambda_{x}\right)|_{\mathrm{s}}.
\end{equation}
On the other hand, $\dot{u}|_{\mathrm{s}}=du_{\mathrm{s}}/dt=2\dot{D}$
by the Rankine\textendash Hugoniot condition \eqref{eq:u_s}, and
therefore we find that the shock acceleration is given by
\begin{equation}
\dot{D}=\frac{1}{2}\left.\left((D-u)u_{x}+\frac{\sigma}{D}\omega\right)\right|_{\shk}.\label{eq:shock-evolution-equation}
\end{equation}
Here, we used the fact that no chemical reaction occurs in the shock,
i.e., $\lambda_{\shk}=0$ at all times, and therefore $0=\dot{\lambda}|_{\mathrm{s}}=\left(\lambda_{t}+D\lambda_{x}\right)|_{\shk}=\left(\omega+D\lambda_{x}\right)|_{\shk}$,
and hence 
\begin{equation}
\lambda_{x}|_{\mathrm{s}}=-\omega_{\shk}/D.\label{eq:lambda_x_s}
\end{equation}

Using \eqref{eq:shock-evolution-equation}, the detonation velocity,
$D\left(t\right)$, can be evolved in time provided the right-hand
side quantities are known. Among these quantities, all are known exactly
in terms of $D$ through the Rankine\textendash Hugoniot conditions,
with the exception of $u_{x}|_{\shk}$. The latter is approximated
using one-sided finite differences for solutions with smooth profiles
of $u$ near the shock. When the solution loses smoothness due to
secondary shocks that can arise behind the lead shock and can catch
up with it, $u_{x}|_{\shk}$ blows up and hence \eqref{eq:shock-evolution-equation}
cannot be applied. The detonation velocity should be computed differently
in that case, which we explain in Subsection \ref{subsec:compute-D(t)}.

\section{Linear stability analysis\label{sec:lin}}

The first step in analyzing the dynamics of traveling-wave solutions
of (\ref{eq:model-Burgers}-\ref{eq:model-reaction}) is to understand
their linear stability. In this section, we investigate the linear
stability properties of the model employing the algorithm developed
in \citep{kabanov2018linear} as well as the traditional method of
normal modes. The neutral stability boundary in the plane of parameters
$q$ and $\theta$ is determined. Note that $q$ and $\theta$ are
the only free parameters of the problem and therefore the neutral
boundary provides a complete stability diagram.

\subsection{Algorithm for linear stability computations\label{sec:lin:algo}}

Let $\vec z=\left(u,\lambda\right)^{\T}$ denote the vector of state
variables. Then, expanding $\vec z$ and $D$ about their steady-state
values, $\vec z=\bar{\vec z}(x)+\vec z'(x,t)$, $D=\bar{D}+\psi'(t)$,
with primes denoting small perturbations, we arrive at the linearized
equations
\begin{equation}
\vec z'_{t}=-\mat A(\bar{\vec z})\vec z'_{x}-\mat B(\bar{\vec z})\vec z'+\frac{d\bar{\vec z}}{dx}\psi',\label{eq:lin:algo:gov-eq}
\end{equation}
where
\[
\mat A=\begin{bmatrix}\bar{u}-\bar{D} & \sigma\\
0 & -\bar{D}
\end{bmatrix},\quad\mat B=\begin{bmatrix}\frac{d\bar{u}}{dx} & 0\\
-\bar{\omega}_{u} & -\bar{\omega}_{\lambda}
\end{bmatrix},
\]
in which the steady quantities are known with the partial derivatives
of $\bar{\omega}$ being $\bar{\omega}_{u}=\theta\sqrt{q}\bar{\omega}$,
$\bar{\omega}_{\lambda}=k\exp\left(\theta\left(\sqrt{q}\bar{u}+q\bar{\lambda}\right)\right)\,\left(\theta q(1-\bar{\lambda})-1\right)$,
and the perturbations $u'$, $\lambda'$, and $\psi'$ are to be found.

Linearization of the Rankine\textendash Hugoniot conditions (\ref{eq:lambda_s}\textendash \ref{eq:u_s})
gives
\begin{equation}
\lambda'_{\mathrm{s}}=0,\quad u'_{\mathrm{s}}=2\psi',\label{eq:fsm:linearization:rh:conditions}
\end{equation}
and linearization of the shock-evolution equation \eqref{eq:shock-evolution-equation}
gives
\begin{equation}
\frac{d\psi'}{dt}=\frac{1}{2}\left[\frac{qk\exp(2\theta\sqrt{q}\bar{D})(\theta\sqrt{q}\bar{D}-1)}{\bar{D}^{2}}\psi'-\bar{D}u'_{x}|_{\mathrm{s}}\right].\label{eq:fsm:linearization:sheveq:fsm:sheveq}
\end{equation}
As mentioned earlier, the shock-evolution equation contains the unknown
perturbation gradient, $u'_{x}|_{\mathrm{s}}$, which must be approximated
numerically using the known values of $u'$ near the shock.

We now explain the algorithm that is used to integrate the linearized
system \citep{kabanov2018linear}. Computations start with the evaluation
of the following parameters: the value of self-sustained detonation
velocity, $\bar{D}$, value of the pre-exponential factor, $k$, and
the numerical reaction-zone length, $\mathcal{L}$, which is found
by integrating \eqref{eq:znd:dlambda_dx} up to $\bar{\lambda}=\bar{\lambda}^{*}$:
\begin{equation}
\mathcal{L}\left(\bar{\lambda}^{*}\right)=\lceil I\rceil\text{ with }I=\int_{0}^{\bar{\lambda}^{*}}\frac{\bar{D}}{\bar{\omega}\left(\bar{u},\bar{\lambda}\right)}\,d\bar{\lambda},\label{eq:algo:znd:partial-length}
\end{equation}
where $\lceil\cdot\rceil$ is the ceiling function, $\bar{u}=\bar{u}\left(\bar{\lambda}\right)$
and $\bar{\omega}\left(\bar{u},\bar{\lambda}\right)$ are given by
\eqref{eq:znd:u-of-lambda} and \eqref{eq:model:reaction-rate}, and
$\bar{\lambda}^{*}=1-\tau_{\lambda}$ with $\tau_{\lambda}$ being
a prescribed tolerance measuring the deviation of $\bar{\lambda}^{*}$
from the chemical-equilibrium value $\bar{\lambda}=1$ to avoid the
divergence of $I$ (to remind, $\bar{\omega}\sim1-\bar{\lambda}$,
which leads to a logarithmic divergence of the integral as $\bar{\lambda}^{*}\to1$).
For the following computations, we use $\tau_{\lambda}=10^{-6}$.

\begin{figure}
\begin{centering}
\includegraphics{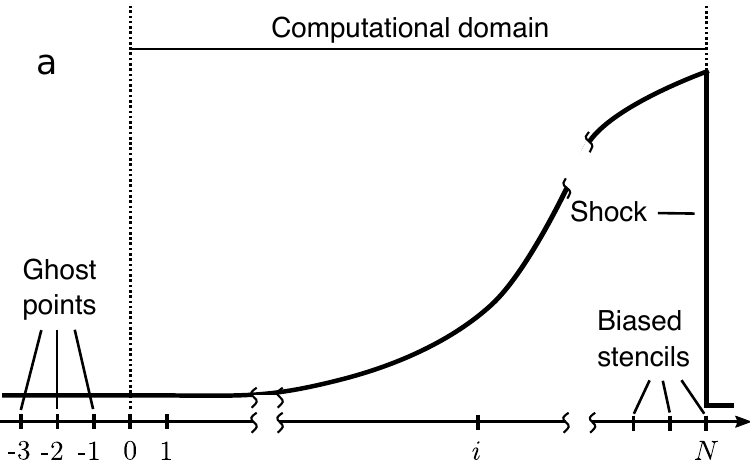} \hfill{}\includegraphics{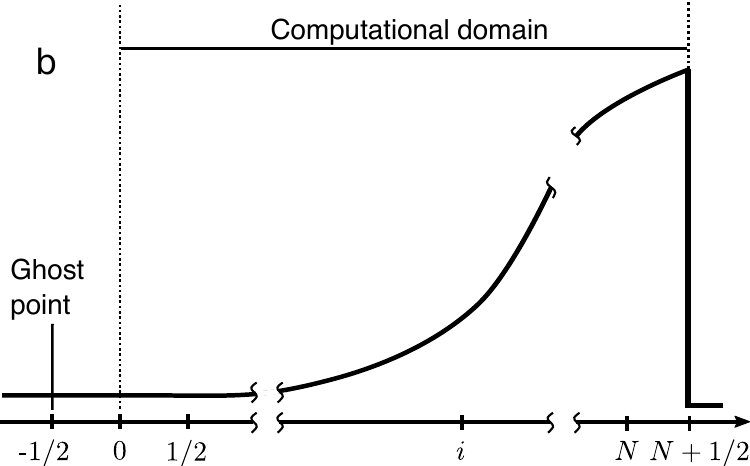}
\par\end{centering}
\caption{The sketches of the numerical grids for: a) linear solver; b) nonlinear
solver (see Section \ref{sec:nonlinear}). The sketched profile is
$u$ as in a traveling-wave solution.\label{fig:algo:grid}}
\end{figure}

The computational domain is partitioned using a uniform grid of size
$N$ with the grid size $\Delta x=1/\Nhrz$, where $\Nhrz$ is the
resolution per half-reaction zone (recall that by \eqref{eq:znd:k},
the half-reaction zone is unity), such that $N=\Nhrz\mathcal{L}$.
Figure \ref{fig:algo:grid}a shows the schematics of the computational
domain and the grid organization for the linear solver while Figure
\ref{fig:algo:grid}b shows the grid for the nonlinear solver, which
is described in Subsection \ref{subsec:nonlinear:desc}.

Once the grid partitioning is completed, the steady-state solution
profiles are found by solving an initial-value problem for the ODE
\eqref{eq:znd:dlambda_dx} with initial conditions at the shock given
by the Rankine\textendash Hugoniot conditions. We use the VODE solver
\citep{brown1989vode} from \texttt{scipy.integrate} package~\citep{scipy}
with a BDF (backward differencing formula) method of fifth order and
relative and absolute tolerances set to $10^{-15}$. After the integration,
all other steady-state quantities ($\bar{u}$, $\bar{\omega}$, $\bar{\omega}_{u}$,
$\bar{\omega}_{\lambda}$, $d\bar{u}/dx$, and $d\bar{\lambda}/dx$)
are computed.

Then, the linearized system is integrated by the method of lines:
\begin{align}
\frac{d\vec z'}{dt} & =\widehat{\vec L}(\bar{\vec z},\vec z'),\quad\text{for }i=0,\quad N-1,\label{eq:lin:algo:mol-a}\\
\frac{d\psi'}{dt} & =\widehat{s}\left(\bar{\vec z},\vec z'\right),\label{eq:lin:algo:mol-b}
\end{align}
where $\widehat{\vec L}(\bar{\vec z},\vec z')$ and $\widehat{s}(\bar{\vec z},\vec z')$
are the finite-difference analogs of the right-hand sides of \eqref{eq:lin:algo:gov-eq}
and \eqref{eq:fsm:linearization:sheveq:fsm:sheveq} in which spatial
derivatives $\vec z'_{x}$ are approximated using finite-difference
formulas given next.

At points $i=0,\dots,N-3$, we compute left- and right-biased approximations
of $\vec z'_{x}$ using the fifth-order upwind method:
\begin{align}
\left.\vec z_{x}'^{-}\right|_{x=x_{i}} & =\frac{-2\vec z'_{i-3}+15\vec z'_{i-2}-60\vec z'_{i-1}+20\vec z'_{i}+30\vec z'_{i+1}-3\vec z'_{i+2}}{60\DeltaX}+\bigO\left(\DeltaX^{5}\right),\label{eq:lin:algo::upwind5-left}\\
\left.\vec z_{x}'^{+}\right|_{x=x_{i}} & =\frac{3\vec z'_{i-2}-30\vec z'_{i-1}-20\vec z'_{i}+60\vec z'_{i+1}-15\vec z'_{i+2}+2\vec z'_{i+3}}{60\DeltaX}+\bigO\left(\DeltaX^{5}\right),\label{eq:lin:algo:upwind5-right}
\end{align}

To avoid the approximation of the spatial derivatives across the shock,
at points $i=\{N-2,N-1,N\}$ we use biased stencils for finite-difference
approximations \citep{HenrickAslamPowers2006}:

\begin{equation}
\left.\vec z'_{x}\right|_{x=x_{N-2}}=\frac{-2\vec z'_{N-5}+15\vec z'_{N-4}-60\vec z'_{N-3}+20\vec z'_{N-2}+30\vec z'_{N-1}-3\vec z'_{N}}{60\DeltaX}+\bigO\left(\Delta x^{5}\right),
\end{equation}

\begin{equation}
\vec z'_{x}|_{x=x_{N-1}}=\frac{-\vec z'_{N-4}+6\vec z'_{N-3}-18\vec z'_{N-2}+10\vec z'_{N-1}+3\vec z'_{N}}{12\DeltaX}+\bigO\left(\Delta x^{4}\right),
\end{equation}

\begin{equation}
u'_{x}|_{N}=\frac{-12u'_{N-5}+75u'_{N-4}-200u'_{N-3}+300u'_{N-2}-300u'_{N-1}+137u'_{N}}{60\DeltaX}+\bigO\left(\Delta x^{5}\right).
\end{equation}
Note that at $x=x_{N}$, only an approximation of $u'_{x}$ is needed.
Once the spatial derivatives are approximated, the right-hand sides
in (\ref{eq:lin:algo:mol-a}\textendash \ref{eq:lin:algo:mol-b})
are computed using the Lax\textendash Friedrichs flux:
\begin{equation}
\widehat{\vec L}(\bar{\vec z},\vec z'_{x})=\vec L\left(\bar{\vec z},\frac{\vec z_{x}'^{+}+\vec z_{x}'^{-}}{2}\right)-\alpha\frac{\vec z_{x}'^{+}-\vec z_{x}'^{-}}{2}.
\end{equation}

After evaluation of the right-hand sides, the system is integrated
in time using the adaptive Runge\textendash Kutta method of order
5(4) due to Dormand and Prince \citep{hairer1993solving}. During
the integration, we record the evolution of the perturbation of detonation
velocity, $\psi'$, by sampling the solution with constant time step,
$\DeltaT=0.005$.

Initial conditions are computed by specifying the initial amplitude
of the perturbation, $A_{0}$, using formulas
\begin{equation}
u'_{0}(x)=2A_{0}\frac{\bar{u}(x)}{\bar{u}_{\shk}},\quad\lambda'_{0}(x)=A_{0}\bar{\lambda}(x),\quad\psi'_{0}=A_{0},
\end{equation}
such that the perturbation is a small multiple of the steady-state
solution. Usually, we use $A_{0}=10^{-10}$ in the following computations.

Boundary conditions must be specified for $u'$ and $\lambda'$. At
the upstream boundary, the following Dirichlet conditions are used
\begin{equation}
u_{N}'=2\psi',\quad\lambda_{N}'=0,\label{eq:fsm:numerical-method:bcs:upstream}
\end{equation}
which are simply the Rankine\textendash Hugoniot conditions \eqref{eq:fsm:linearization:rh:conditions}.

At the downstream boundary, the zeroth-order extrapolation is used
\begin{equation}
u_{i}'=u_{0}',\quad\lambda_{i}'=\lambda_{0}'\quad\mbox{for }i=\{-3,-2,-1\}.\label{eq:fsm:numerical-method:bcs:downstream}
\end{equation}
Strictly speaking, such an extrapolation results in some reflections
from the boundary. However, the forward-going characteristics that
transfer information from this boundary to the shock are almost vertical,
and therefore, do not affect the shock evolution over the times of
integration \citep{kasimov2004dynamics}.

The computed solutions are analyzed using a postprocessing algorithm
based on the Dynamic Mode Decomposition (DMD) \citep{schmid2010dynamic}.
In this algorithm, the time snapshots of the system state, $\vec x_{i}\in\R^{m}$,
$i=1,\dots,n$,\textbf{ }are stacked into matrices
\[
\mat X=[\vec x_{0},\vec x_{1},\dots,\vec x_{n-1}],\quad\mat Y=[\vec x_{1},\vec x_{2},\dots,\vec x_{n}],
\]
and we search for a linear mapping $\mat A\in\R^{m\times m}$ such
that $\mat Y=\mat A\mat X$. Formally, $\mat A=\mat Y\mat X^{\dagger}$,
where $\dagger$ denotes the Moore-Penrose pseudoinverse. However,
we are not interested in the mapping $\mat A$ \emph{per se,} but
in its most significant eigenvalues which determine the dynamics of
the observed system. The algorithm consists of the following steps:
\begin{enumerate}
\item Compute the reduced singular value decomposition \citep{trefethen1997numerical}
of $\mat X$: $\mat X=\mat U\mat{\Sigma}\mat V^{\T}$, where $\mat U$
and $\mat V$ are matrices with orthonormal columns and $\mat{\Sigma}=\diag(\sigma_{1},\dots,\sigma_{\min(m,n)})$.
\item Find $\mat U_{r}$, $\mat{\Sigma}_{r}$, and $\mat V_{r}$ by truncating
$\mat U$, $\mat{\Sigma}$, and $\mat V$ to some reasonably small
rank $r$. The algorithm of determining the rank $r$ is described
below.
\item Define an additional matrix $\widetilde{\mat A}=\mat U_{r}^{\T}\mat A\mat U_{r}=\mat U_{r}^{\T}\mat Y\mat V_{r}\mat{\Sigma}_{r}^{-1}$,
where $\widetilde{\mat A}\in\R^{r\times r}$.
\item Compute the eigenvalues and the eigenvectors of $\widetilde{\mat A}$:
$\widetilde{\mat A}\mat W=\mat W\mat{\Lambda}$, where $\mat{\Lambda}=\diag\left(\lambda_{1},\dots,\lambda_{r}\right)$,
and $\mat W\in\R^{r\times r}$ is the matrix whose columns are the
eigenvectors of $\widetilde{\mat A}$.
\item The eigenvalues of $\widetilde{\mat A}$ are also the eigenvalues
of $\mat A$. The corresponding eigenvectors of $\mat A$ are found
from $\mat{\Phi}=\mat Y\mat V_{r}\mat{\Sigma}_{r}^{-1}\mat W\mat{\Lambda}^{-1}$
\citep{tu2014dynamic}.
\end{enumerate}
The pairs $\left(\lambda_{i},\vec{\phi}_{i}\right)$, where $\vec{\phi}_{i}$
is the $i$-th column of $\mat{\Phi}$, constitute the DMD modes.
The eigenvalue $\lambda_{i}$ with $|\Re\left(\lambda_{i}\right)|\leq1$
implies stability of the $i$-th mode. We convert these eigenvalues
to the continuous-time eigenvalues $\alpha_{i}$:
\begin{equation}
\alpha_{i}=\frac{\log\lambda_{i}}{\DeltaT},\quad i=1,\dots,r,
\end{equation}
where $\Re\left(\alpha_{i}\right)\leq0$ implies stability of the
$i$-th mode.

Instead of constructing matrices $\mat X$ and $\mat Y$ using $m$-dimensional
system state, we construct a Hankel matrix $\mat Z\in\R^{L\times\left(n-L+1\right)}$
from the one-dimensional time series of the perturbation of detonation
velocity $\psi'$ with $L=1000$. Then $\mat X$ and $\mat Y$ are
constructed from $\mat Z$ by excluding the last and first columns
of $\mat Z$, respectively.

Now we explain the algorithm for determining the rank $r$. First,
we find the largest possible rank $R$ from the condition $\sigma_{R+1}/\sigma_{1}<10^{-10}$.
Second, we construct the list of the possible ranks by considering
all $i=1,\dots,R$, and if for some $i$ it follows that $\sigma_{i+1}/\sigma_{i}\leq0.95$,
then $i$ is added to the list. The rationale for this is that the
sufficient gap between $\sigma_{i}$ and $\sigma_{i+1}$ signals that
they do not correspond to the conjugate pair of eigenvalues and hence
$i$ is a possible rank. Once the list of the possible ranks is constructed,
for each rank in the list we compute the dynamic mode decomposition
by the algorithm described above and then compute the corresponding
fit and residual errors:
\begin{equation}
e_{\text{fit}}=\frac{\|\widehat{\psi}'-\psi'\|_{2}}{\|\psi'\|_{2}},\quad e_{\text{resid}}=\|\mat Y-\mat{\Phi}\mat{\Lambda}\mat{\Phi}^{\dagger}\mat X\|_{2},
\end{equation}
where $\widehat{\psi}'$ is the DMD reconstruction of $\psi'$. Then
we find the decompositions DMD$_{1}$ and DMD$_{2}$ with the smallest
fit errors $e_{\text{fit},1}$ and $e_{\text{fit},2}$. If $0.5\leq e_{\text{fit},1}/e_{\text{fit},2}\leq1$,
then the corresponding residual errors $e_{\text{resid},1}$ and $e_{\text{resid},2}$
are also considered. If $e_{\text{resid},1}<e_{\text{resid},2}$,
then the best decomposition is DMD$_{1}$, otherwise \textendash{}
DMD$_{2}$. Once the best DMD is determined, we sort the DMD eigenvalues
by imaginary and real parts, removing the eigenvalues that have the
negative imaginary parts or the real parts less than $-1$ (the latter
are considered too stable). Stable eigenvalues $\alpha_{i}$ with
$-1\leq\Re\alpha_{i}\leq0$ are preserved, which is required for the
algorithm of the construction of the neutral stability curve in Subsection
\ref{subsec:neutral-stability}.

\subsection{Results of the linear stability computations\label{sec:results}}

\subsubsection{Time series of detonation velocity and perturbation profiles}

\begin{figure}
\begin{centering}
\includegraphics{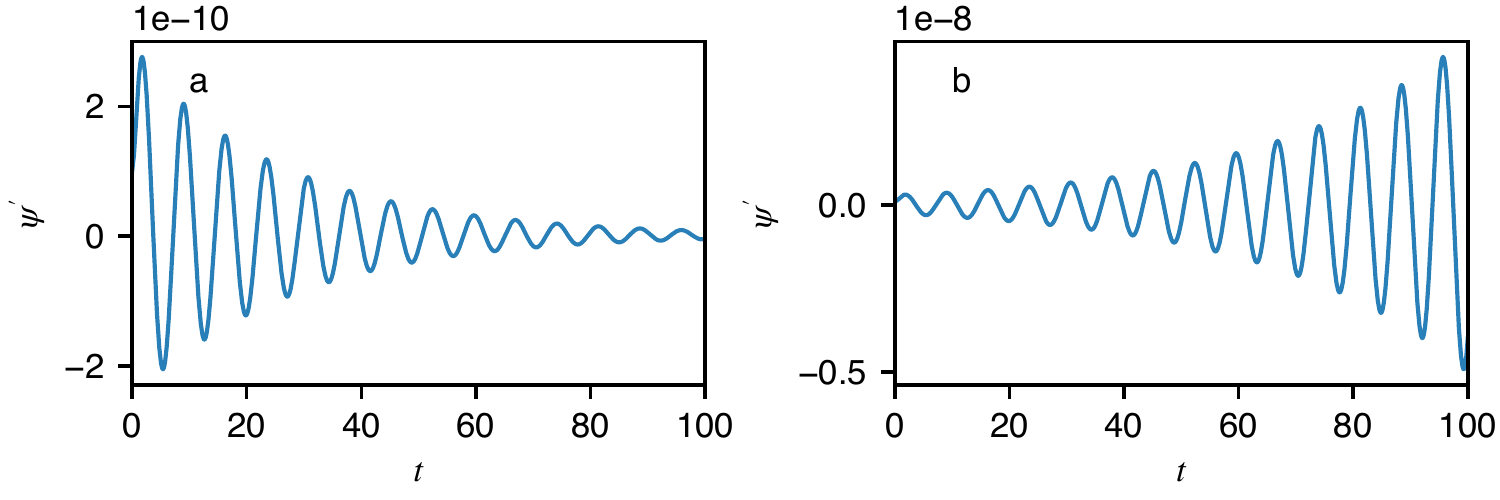}
\par\end{centering}
\caption{Time series of the perturbation of detonation velocity $\psi'$ for
$q=4$: a) stable solution with $\theta=0.92$; b) unstable solution
with $\theta=0.95$.\label{fig:results:time-series}}
\end{figure}

\begin{figure}
\begin{centering}
\includegraphics{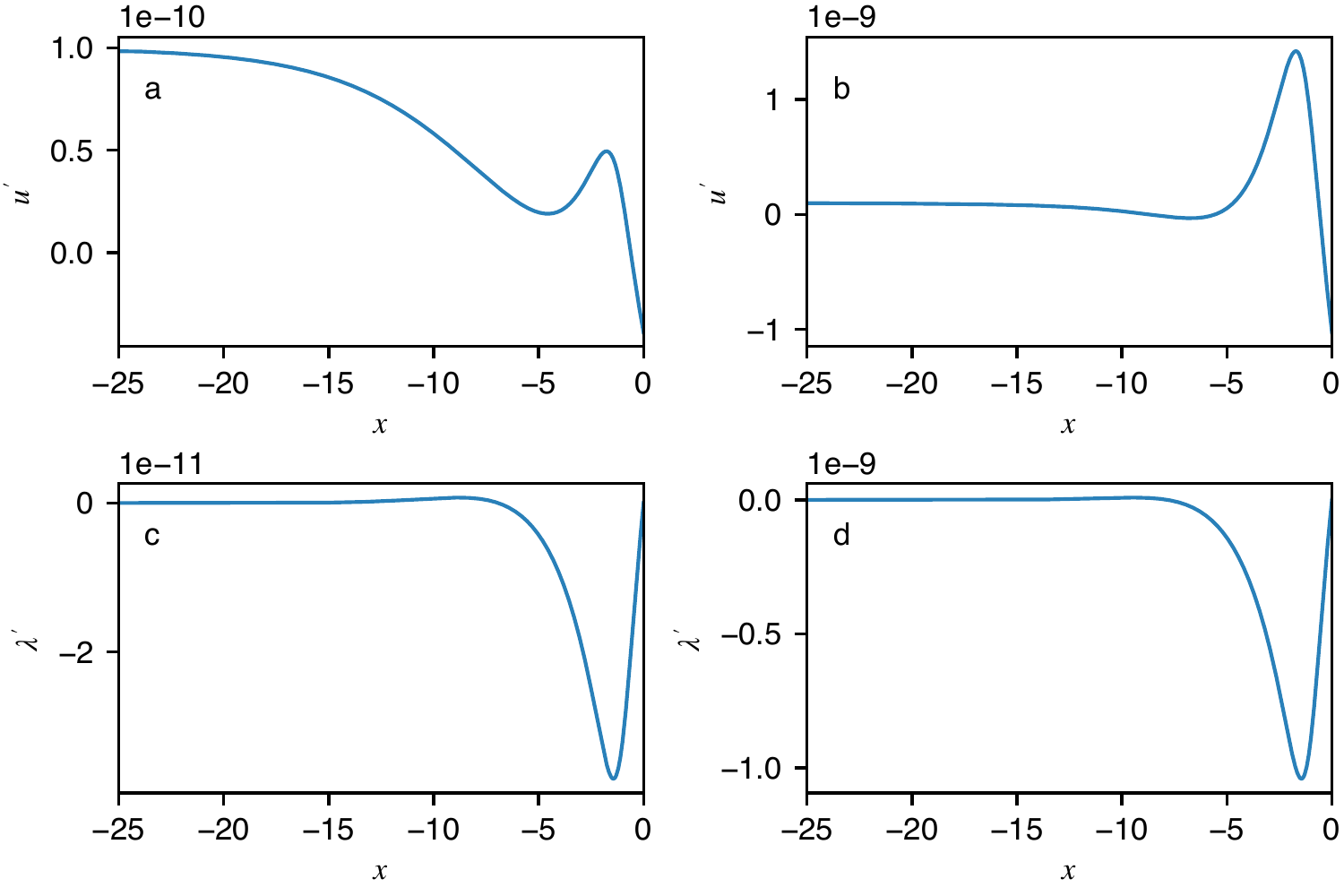}
\par\end{centering}
\caption{Perturbation profiles for $q=4$ recorded at time $t=50$: a-b) perturbations
of velocity $u'$, c-d) perturbations of reaction progress variable
$\lambda'$. Subfigures a) and c) are for $\theta=0.92$ and subfigures
b) and d) are for $\theta=0.95$.\label{fig:results:linear-pert}}
\end{figure}

Figure \ref{fig:results:time-series} shows the computed time series
of the perturbation of detonation velocity, $\psi'$, for stable and
unstable solutions for $q=4$ with $\theta=0.92$ and $\theta=0.95$.
In Figure \ref{fig:results:time-series}a, a stable case is shown
in which initial perturbation of the detonation velocity decays, and
at long times $D\to D_{\mathrm{CJ}}$. In contrast, Figure \ref{fig:results:time-series}b
shows the unstable case, in which the detonation velocity oscillates
around the CJ value with a growing amplitude of the oscillations.
Thus, Figure \ref{fig:results:time-series} indicates that an Andronov-Hopf
bifurcation occurs in the system. Figure \ref{fig:results:linear-pert}
shows the corresponding perturbation profiles of $u'$ and $\lambda'$
that are sampled at time $t=50$. They illustrate the typical behavior
of eigenfunctions, whose dynamics are concentrated in the near-shock
region.

\subsubsection{Comparison with normal-mode analysis}

To check the validity of our computations, we also investigate the
system via the normal-mode analysis \citep{LeeStewart90}. The governing
equations are linearized about the ZND solution and normal-mode expansions
are assumed for the unknowns, $\vec z(x,t)=\bar{\vec z}\left(x\right)+\epsilon\vec z'(x)\exp\left(\alpha t\right)$,
$D(t)=\bar{D}+\epsilon\psi(t),$ where $\left|\epsilon\right|\ll1$,
$\psi(t)=\exp\left(\alpha t\right)$ is proportional to the perturbation
of the unsteady shock relative to its steady-state position, $x(t)=\bar{D}t$,
and $\alpha$ is the complex growth rate which is the eigenvalue of
the problem. A boundary-value problem is then posed for $\vec z'(x)$
with the Rankine\textendash Hugoniot conditions on the right boundary
(on the shock) and a boundedness condition on the left boundary (at
the chemical equilibrium). Substitution of the normal-mode expansions
into the governing equations leads to a linearized system for perturbations
$\vec z'$:
\[
\alpha\vec z'+\mat A(\bar{\vec z})\frac{d\vec z'}{dx}+\mat C\left(\bar{\vec z}\right)\vec z'-\alpha\vec b=0,
\]
where
\[
\mat A(\bar{\vec z})=\begin{bmatrix}\bar{u}-\bar{D} & \sigma\\
0 & -\bar{D}
\end{bmatrix},\quad\mat C\left(\bar{\vec z}\right)=\begin{bmatrix}\frac{d\bar{u}}{dx} & 0\\
-\bar{\omega}_{u} & -\bar{\omega}_{\lambda}
\end{bmatrix},\quad\vec b=\begin{bmatrix}\frac{d\bar{u}}{dx}\\
\frac{d\bar{\lambda}}{dx}
\end{bmatrix}
\]
with $\alpha\in\mathbb{C}$, $\vec z'(x,t)\in\mathbb{C}^{2}$. We
reformulate the problem in terms of the steady-state reaction progress
variable $\bar{\lambda}$ instead of $x$ using \eqref{eq:znd:eqn-of-reaction}
and arrive at the system
\begin{equation}
\frac{d\vec z'}{d\bar{\lambda}}=-\frac{\bar{D}}{\bar{\omega}}\mat A^{-1}\left[-\left(\alpha\mat I+\mat C\right)\vec z'+\alpha\vec b\right]\label{eq:nm-system}
\end{equation}
 which is subject to the linearized Rankine\textendash Hugoniot conditions
at $\bar{\lambda}=0$,
\begin{equation}
u'=2\alpha,\quad\lambda'=0.\label{eq:nm-rh}
\end{equation}

\begin{figure}
\begin{centering}
\includegraphics{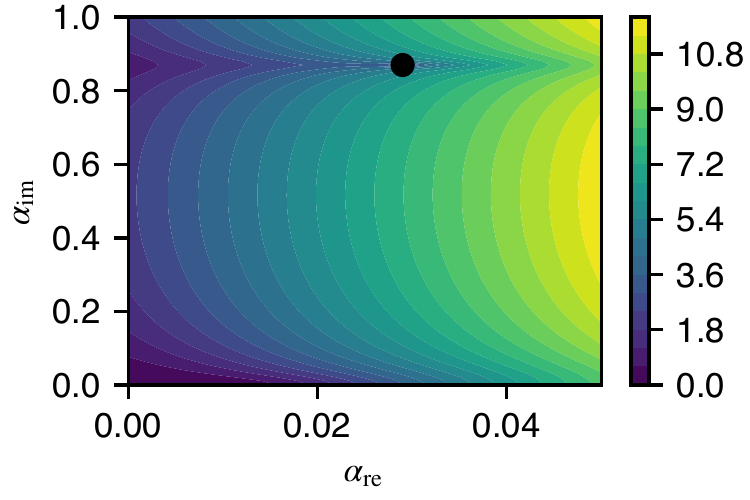}
\par\end{centering}
\caption{The carpet plot of $\log(1+|H|)$, where $H$ is the boundedness function
\eqref{eq:nm-boundedness} for $q=4$, $\theta=0.95$. Dot $\left(0.029,0.87\right)$
denotes an approximate local minimum of $\log(1+|H|)$.\label{fig:Carpet}}
\end{figure}

The boundedness condition at $x\to-\infty$ is required which expresses
the fact that eigenfunctions of the problem must remain bounded at
the end of the reaction zone. In \citep{StewartKasimovSIAP05}, a
strategy is explained for the derivation of the boundedness condition
through the linearization of the forward characteristic equation.
Application of this strategy to our model yields 
\begin{equation}
H(\alpha)=\alpha\left(\bar{u}u'+\sigma\lambda'\right)-\sigma\bar{\omega}_{\lambda}\lambda'=0,\label{eq:nm-boundedness}
\end{equation}
where $H$ denotes the boundedness function. Thus the problem is reduced
to the boundary-value problem (\ref{eq:nm-system}\textendash \ref{eq:nm-boundedness})
which has a bounded solution only for particular values of $\alpha$,
to be determined. The strategy of solving the problem is based on
the shooting method. Namely, (\ref{eq:nm-system}\textendash \ref{eq:nm-rh})
is solved starting at the shock and integrating toward the end of
the reaction zone, where condition \eqref{eq:nm-boundedness} must
be satisfied. Solving multiple initial-value problems for a range
of $\alpha$ allows to plot a ``carpet'' of $\log(1+|H|)$, in which
local minima can be identified that correspond to approximate locations
of $\alpha$ that satisfy the boundary-value problem (\ref{eq:nm-system}\textendash \ref{eq:nm-boundedness}).
Figure \ref{fig:Carpet} shows such a plot for $q=4$, $\theta=0.95$
for the range $\alpha_{\text{re}}\in\{0,0.001,0.002,\dots,0.05\}$,
$\alpha_{\text{im}}\in\{0,0.01,0.02,\dots,1]$ with the only local
minimum found, which is indicated by a red dot with coordinates $(0.029,0.87)$.

Subsequently, using the approximated value of $\alpha$ as an initial
guess, we employ a root solver to accurately compute $\alpha$ that
satisfies \eqref{eq:nm-boundedness}. For this, \emph{fsolve} routine
of \emph{scipy} package is used \citep{scipy}. The results are shown
in Table \ref{tab:results:nm-comparison} along with the results found
with the method described in Subsection \ref{sec:lin:algo}. Excellent
agreement between the two approaches is seen, as growth rates $\alpha_{\text{re}}$
and frequencies $\alpha_{\text{im}}$ match to four and six significant
digits, respectively.

\begin{table}
\begin{centering}
\begin{tabular}{lcc}
\toprule 
Approach & $\alpha_{\text{re}}$ & $\alpha_{\text{im}}$\tabularnewline
\midrule
Linear unsteady analysis & 0.02909342 & 0.87041209\tabularnewline
Normal-mode analysis & 0.02909286 & 0.87041272\tabularnewline
\bottomrule
\end{tabular}
\par\end{centering}
\caption{Comparison of the growth rates $\alpha_{\text{re}}$ and frequencies
$\alpha_{\text{im}}$ of perturbations obtained via linear unsteady
computations and normal-mode computations for $q=4$, $\theta=0.95$.\label{tab:results:nm-comparison}}
\end{table}

\subsubsection{Migration of the linear spectrum as activation energy is varied}

\begin{figure}
\begin{centering}
\includegraphics{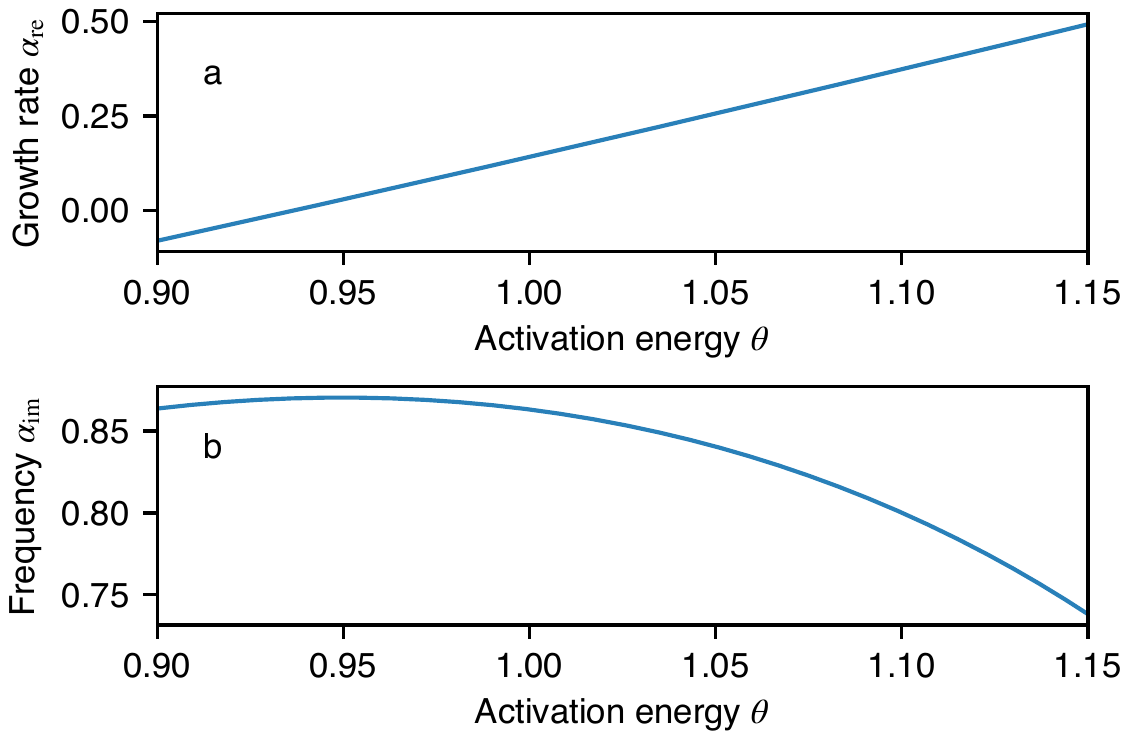}
\par\end{centering}
\caption{Migration of the linear spectrum for $q=4$ as activation energy $\theta$
is varied: a) growth rate, b) frequency. Each curve consists of 251
points. Only one mode in the spectrum was found.\label{fig:linear-spectrum}}
\end{figure}

Figure \ref{fig:linear-spectrum} shows how the linear spectrum changes
as the activation energy, $\theta$, increases in the range $[0.90;1.15]$
with step $\Delta\theta=0.001$. Only one mode in the spectrum is
found. Its growth rate increases from $-0.081$ to $0.493$ almost
linearly with $\theta$ and a bifurcation to instability occurs at
$\theta_{\text{crit}}=0.937\pm0.001$. The frequency of the mode exhibits
slightly more complicated behavior: it initially increases from 0.864
at $\theta=0.90$, reaches the maximum value $0.870$ at $\theta=0.95$,
then decreases to $0.738$ at $\theta=1.15$.

\subsubsection{Neutral stability\label{subsec:neutral-stability}}

Now we turn our attention to the determination of the neutral stability
boundary for a wide range of parameters $\theta$ and $q$, that is,
the curve in a $\left(\theta,q\right)$ plane that separates stable
steady-state solutions from unstable ones. The boundary is determined
numerically. For this purpose, we generate 256 linearly spaced values
of $q$ in the range $[0.81,16]$ (corresponds to $\bar{D}$ in the
range $[0.9,4]$) and for each $q$, we find the critical value of
$\theta$ such that the growth rate of instability is close to zero
with a tolerance $10^{-3}$. Finding $\theta_{\text{crit}}$ is based
on the idea of bisection where a range of $\theta$ is recursively
divided in subranges unless the above condition on the growth rate
is satisfied. Initial interval of search for $\theta$ is taken $[0.2;5]$.
Grid resolution used here is $\Nhrz=40$.

\begin{figure}
\centering{}\includegraphics{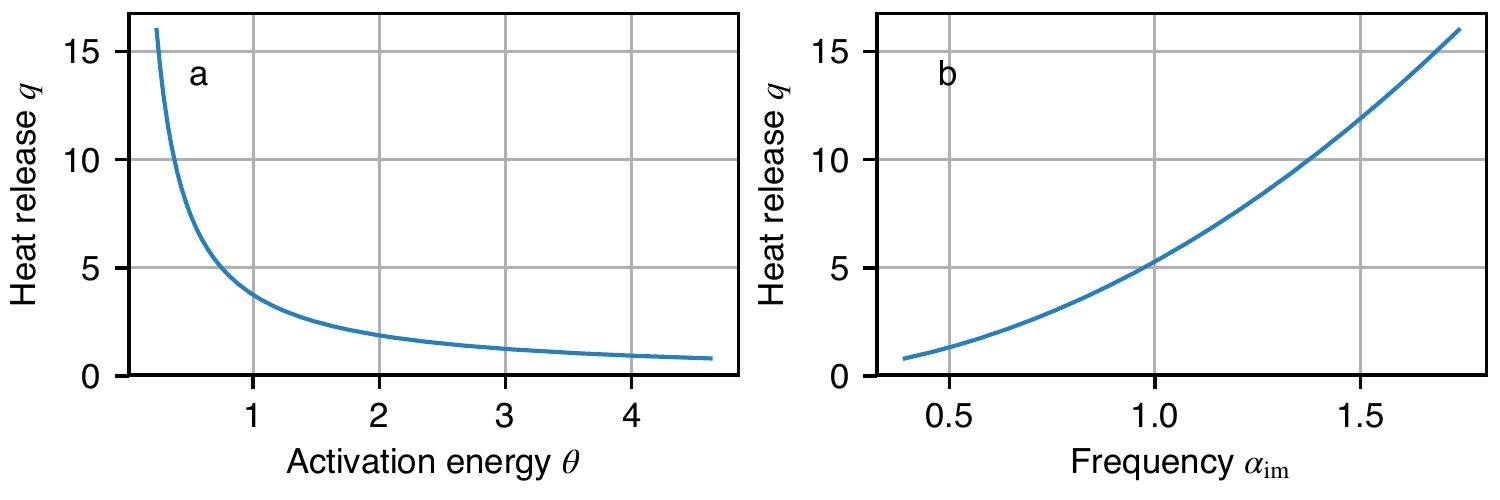}\caption{Neutral stability for the Fickett model: a) neutral stability curve
in $(\theta,q)$ plane; b) frequency of oscillation $\alpha_{\text{im}}$
along the neutral stability curve. Each curve consists of 256 uniformly
spaced points for $q\in[0.81;16]$.\label{fig:neutral-stability}}
\end{figure}

Figure \ref{fig:neutral-stability} shows the computed neutral stability
curve in the $\left(\theta,q\right)$ plane and the frequency of oscillation
$\alpha_{\text{im}}$ along the curve. The data demonstrate that
when $q$ increases, $\theta_{\text{crit}}$ decreases while $\alpha_{\text{im}}$
increases. For the lowest $q$ considered here ($q=0.81$), $\theta_{\text{crit}}=4.625$
and $\alpha_{\text{im}}=0.391$, and for the largest $q$ ($q=16$),
$\theta_{\text{crit}}\approx0.234$ and $\alpha_{\text{im}}=1.74$.

We also provide information about the critical values of activation
energy $\theta_{\text{crit}}$ and the frequency of oscillation $\alpha_{\text{im}}$
on the neutral stability boundary for several values of $q$ in Table
\ref{tab:results:neutral-stability}. It is interesting to note that
the $(\theta,q)$ neutral curve is nearly a hyperbola, $q\theta\approx3.75$.

\begin{table}
\centering{}\caption{Critical values of activation energy $\theta_{\text{crit}}$ and the
frequency of oscillation $\alpha_{\text{im}}$ on the neutral stability
boundary for several values of heat release $q$.\label{tab:results:neutral-stability}}
\begin{tabular*}{\textwidth}{@{\extracolsep{\fill}}cccc} \\
\toprule
$i$ & $q$ & $\theta_\text{crit}$ & $\alpha_\text{im}$ \\
\midrule
1 & \num{0.81} & 4.625 & 0.391 \\
2 & \num{1.00} & 3.746 & 0.435 \\
3 & \num{2.00} & 1.873 & 0.615 \\
4 & \num{4.00} & 0.937 & 0.870 \\
5 & \num{9.00} & 0.417 & 1.305 \\
6 & \num{16.00} & 0.234 & 1.740 \\
\bottomrule
\end{tabular*}

\end{table}

\section{Nonlinear dynamics \label{sec:nonlinear}}

\subsection{Description of the numerical algorithm\label{subsec:nonlinear:desc}}

Having studied the linear stability properties of the traveling-wave
solution and having identified the transition to instability when
$\theta q$ is large enough (larger than approximately $3.75$), we
now turn attention to the question of the nonlinear dynamics of solutions
as we move away from the neutral curve into the unstable domain. To
solve the nonlinear system numerically, we use a second-order MUSCL
scheme with the minmod flux limiter \citep{leveque2002finite}, which
we describe below along with the algorithm for the solution of the
Riemann problem required for this method.

The MUSCL scheme is a conservative Godunov-type method for a hyperbolic
system:
\begin{equation}
\vec z_{t}+\vec f(\vec z)_{x}=\vec s(\vec z)
\end{equation}
with $\vec z\in\R^{n}$ being the vector of unknowns,$\vec f:\R^{n}\to\R^{n}$
the flux function, and $\vec s$ the source term. Partitioning a computational
domain into cells $[x_{i-1/2},x_{i+1/2}]$ (the sketch of the grid
is shown in Figure \ref{fig:algo:grid}b) we obtain a semi-discretized
scheme:
\begin{equation}
\frac{d\bar{\vec z}_{i}}{dt}=-\frac{\DeltaT}{\DeltaX}\left(\vec f_{i+1/2}-\vec f_{i-1/2}\right)+\vec s(\bar{\vec z}_{i}),
\end{equation}
where $\bar{\vec z}_{i}$ is the average value of $\vec z$ for the
computational cell centered at $x_{i}$, and the fluxes at the cell
boundaries $\vec f_{i+1/2}$ are approximated by reconstructing $\vec z_{i+1/2}$.
The reconstruction is done through the solution of the Riemann problem
with initial conditions
\begin{equation}
\vec z_{i+1/2}=\begin{cases}
\vec z_{\text{L}} & x<x_{i+1/2},\\
\vec z_{\text{R}}, & x>x_{i+1/2},
\end{cases}
\end{equation}
where $\vec z_{\text{L}}$ and $\vec z_{\text{R}}$ are given immediately
on the left and the right of $x_{i+1/2}$, respectively. For a second-order
scheme, they are found by assuming linear distribution of $\vec z_{i}(x)$
in the cell $[x_{i-1/2};x_{i+1/2}]$:
\begin{equation}
\vec z_{i}(x)=\bar{\vec z}_{i}+\vec{\sigma}_{i}(x-x_{i}),
\end{equation}
where $\vec{\sigma}_{i}$\textbf{ }is the slope of the linear reconstruction,
therefore,
\begin{align}
\vec z_{\text{L}} & =\bar{\vec z}_{i}+\vec{\sigma}_{i}\frac{\DeltaX}{2},\\
\vec z_{\text{R}} & =\bar{\vec z}_{i+1}-\vec{\sigma}_{i+1}\frac{\DeltaX}{2}.
\end{align}
For the second-order MUSCL schemes, $\vec{\sigma}_{i}=\vec{\sigma}_{i}\left(\bar{\vec z}_{i}-\bar{\vec z}_{i-1},\bar{\vec z}_{i+1}-\bar{\vec z}_{i}\right)$.
We compute $\vec{\sigma}_{i}$ using the minmod flux limiter \citep{leveque2002finite}
given componentwise as: 
\begin{equation}
\vec{\sigma}_{i}(a,b)=\begin{cases}
\min(a,b)/\DeltaX, & \text{if }a>0,b>0,\\
\max(a,b)/\DeltaX, & \text{if }a<0,b<0,\\
0, & \text{if }ab<0.
\end{cases}
\end{equation}

For the model under consideration, solution $\vec z_{i+1/2}$ of the
Riemann problem corresponds to the state $\left(u_{i+1/2},\lambda_{i+1/2}\right)^{\T}$,
which is found by the analysis of the waves that can occur in the
Riemann problem and is given in the next subsection. This solution
propagates along the line $\left(x_{i+1/2},t\right)^{\T}$ in the
$\left(x,t\right)$ plane. As each Riemann problem assumes $x_{i+1/2}=0$
in a local reference frame, the solution is time-independent.

Once all the local Riemann problems are solved over the whole grid,
fluxes $\vec f_{i+1/2}$ at the edges $x_{i+1/2}$ can be computed.
On the left part of the domain, we introduce a ghost point with extrapolation
$u_{-1}=u_{0}$ such that the left-most flux $\vec f_{-1/2}$ can
be computed. On the right boundary the flux $\vec f_{N+1/2}$ is computed
by means of the Rankine\textendash Hugoniot conditions $u_{N+1/2}=2D$,
$\lambda_{N+1/2}=0$, which leads to the fluxes $\vec f_{N+1/2}=0$
both for $u$ and $\lambda$ for the chosen ambient conditions.

The time integrator is an adaptive-step Runge\textendash Kutta integrator
DOPRI5 \citep{hairer1993solving} as used for the linear unsteady
simulations above. It may appear that strong-stability-preserving
Runge\textendash Kutta integrators \citep{gottlieb2001strong} are
more appropriate for the problem at hand as we expect internal shocks
to be generated inside the reaction zone. However, for the reactive
flow, error estimation and consequently adaptive choice of a time
step are crucial components of a time integrator. One cannot rely
on the Courant\textendash Friedrichs\textendash Lewy condition \citep{leveque2002finite}
in the computations of the time step due to chemical reactions occurring
at much smaller time scales than wave propagation. The presence of
the adaptive-step DOPRI5 integrator in the numerical environment that
we use (\emph{scipy} library of the scientific Python stack \citep{scipy})
determined our choice of the time integrator. Relative and absolute
tolerances of the DOPRI5 integrator are set to $10^{-8}$ and $10^{-6}$,
respectively, for the nonlinear simulations.

\subsection{The Riemann problem\label{subsec:riemann}}

Below we provide the solution of the Riemann problem for the model
system. To analyze possible wave configurations, we write the (nonreactive)
system in the characteristic form:
\begin{alignat}{2}
\dot{p} & =0 & \quad\text{on }\dot{x} & =u-D,\\
\dot{\lambda} & =0 & \quad\text{on }\dot{x} & =-D,
\end{alignat}
and from the characteristic equations themselves we can see that the
first wave is nonlinear as the characteristic speed depends on $u$
(and correspondingly on $\lambda$ due to coupling), while the second
wave is linear and, therefore, second wave is always a left-going
wave with speed $-D$. Now, assuming that $u>0$, it is clear that
the first wave is always to the right of the second wave and hence
the solution consists of three distinct regions: L-region to the left
of the second wave, M-region between the second and the first wave,
and R-region to the right of the first wave. Now we consider various
possible configurations.

Case 1. $u_{\text{L}}<u_{\text{R}}$, $\lambda_{\text{L}}<\lambda_{\text{R}}$.
In this case, from the second characteristic equation it follows that
$\lambda$ experiences a jump along the line $x=-Dt$, hence this
line is the trajectory of the contact discontinuity moving to the
left. Then, for M-region, which is to the right of the contact, we
have $\lambda_{\text{M}}=\lambda_{\text{R}}$. The invariant $p$
is conserved along the curves $\dot{x}=u-D$, therefore (due to the
jump in $\lambda$) $u$ must jump on the contact:
\begin{equation}
\frac{1}{2}\left(u_{\text{M}}^{2}+q\lambda_{\text{M}}\right)=\frac{1}{2}\left(u_{\text{L}}^{2}+q\lambda_{\text{L}}\right),
\end{equation}
from which it follows that
\begin{equation}
u_{\text{M}}=\sqrt{u_{\text{L}}^{2}+q(\lambda_{\text{L}}-\lambda_{\text{M}})}=\sqrt{u_{\text{L}}^{2}+q(\lambda_{\text{L}}-\lambda_{\text{R}})}\label{eq:rp-u_m}
\end{equation}
and hence $u_{\text{M}}<u_{\text{L}}<u_{\text{R}}$. Therefore, the
first wave is a centered rarefaction wave, with constraining characteristics
\begin{equation}
x_{\text{HEAD}}=\left(u_{\text{R}}-D\right)t,\quad x_{\text{TAIL}}=\left(u_{\text{M}}-D\right)t,\label{eq:riemann-rarefaction-speed}
\end{equation}
with the solution inside the rarefaction wave being $u=x/t+D$, which
is found from the integration of $\dot{x}=u-D$ with initial conditions
$x=0$, $t=0$.

Case 2. $u_{\text{L}}\geq u_{\text{R}}$, $\lambda_{\text{L}}\geq\lambda_{\text{R}}$.
In this case $\lambda_{\text{M}}=\lambda_{\text{R}}$ again and $u_{\text{M}}$
is computed using~\eqref{eq:rp-u_m} with $u_{\text{M}}\geq u_{\text{L}}\geq u_{\text{R}}$,
that is, the characteristics of the first family from M- and R-regions
collide and form a shock wave that propagates with speed
\begin{equation}
s=\frac{u_{\text{R}}+u_{\text{M}}}{2}-D.
\end{equation}

Case 3. $u_{\text{L}}<u_{\text{R}}$, $\lambda_{\text{L}}\geq\lambda_{\text{R}}$.
In this case $\lambda_{\text{M}}=\lambda_{\text{R}}$ and $u_{\text{M}}$
is found using Eq.~\eqref{eq:rp-u_m}. Hence, the first wave is either
a centered rarefaction wave (if $u_{\text{M}}<u_{\text{R}}$) as in
Case~1 or a shock wave (if $u_{\text{M}}\geq u_{\text{R}}$) as in
Case~2 while the second wave is a contact discontinuity.

Case 4. $u_{\text{L}}\geq u_{\text{R}}$, $\lambda_{\text{L}}<\lambda_{\text{R}}$.
In this case $\lambda_{\text{M}}=\lambda_{\text{R}}$ and $u_{\text{M}}$
is found using formula~\eqref{eq:rp-u_m}. Then the first wave is
either a centered rarefaction or a shock wave depending on whether
$u_{\text{M}}<u_{\text{R}}$ or $u_{\text{M}}\geq u_{\text{R}}$,
that is, this case is the same as Case~3.

For the Godunov method, we are interested in the solution of the Riemann
problem along the trajectory $x=0$ only. Summarizing the cases considered
above, we conclude that if $u_{\text{R}}>D$, then $u_{i+1/2}=u_{\text{M}}$,
$\lambda_{i+1/2}=\lambda_{\text{M}}$. However, the flow can become
supersonic with $u_{\text{R}}<D$, particularly near the left boundary,
and then not only contact discontinuity moves to the left, but the
rarefaction or shock wave as well. In these cases, the solution of
the Riemann problem is $u_{i+1/2}=u_{\text{R}}$, $\lambda_{i+1/2}=\lambda_{\text{R}}$.

\begin{figure}
\begin{centering}
\includegraphics{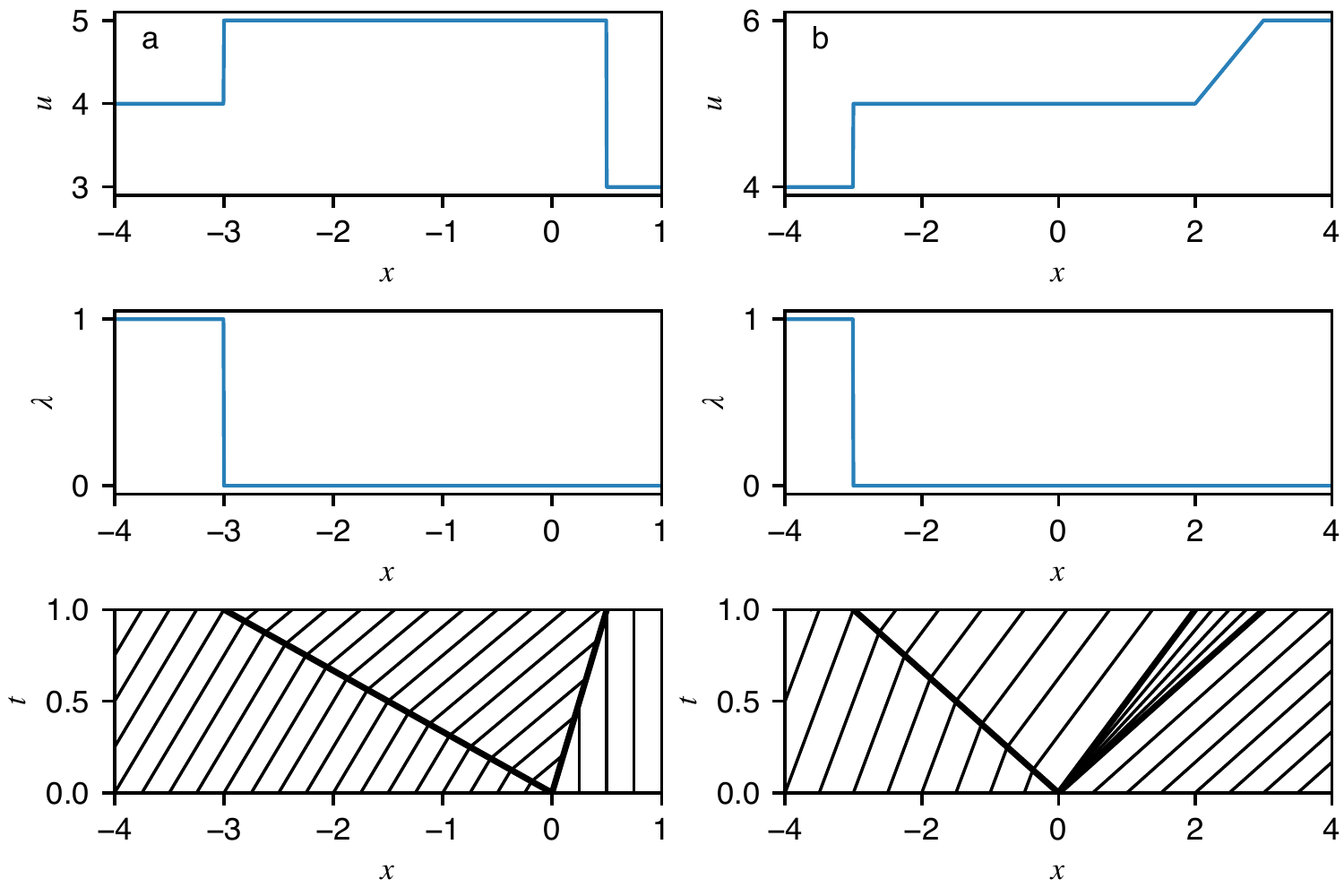}
\par\end{centering}
\caption{Solution of the Riemann problem and characteristics trajectories up
to the final time $t=1$ for $q=9$, $D=3$ and two different initial
conditions: a) $\left(u_{\text{L}},\lambda_{\text{L}}\right)=\left(4,1\right)$,
$\left(u_{\text{R}},\lambda_{\text{R}}\right)=\left(3,0\right)$;
b) $\left(u_{\text{L}},\lambda_{\text{L}}\right)=\left(4,1\right)$,
$\left(u_{\text{R}},\lambda_{\text{R}}\right)=\left(6,0\right)$.
In both a) and b), bottom plots show the characteristics of the first
family, while the characteristics of the second family are not shown
except for the trajectory of the contact discontinuity, $x=-Dt$,
(all other characteristics of this family are parallel to it). \label{fig:solutions-Riemann-problems}}
\end{figure}

To illustrate the complete solutions of the Riemann problem and characteristic
trajectories, we provide two examples. For both problems, we take
$q=9$, hence $D=\sqrt{q}=3$, $\lambda_{\text{L}}=1$, $\lambda_{\text{R}}=0$,
and the final time $t=1$. Figure \ref{fig:solutions-Riemann-problems}a
shows the solution of the Riemann problem for $u_{\text{L}}=4$, $u_{\text{R}}=3$,
along with the characteristics. As can be seen, the solution consists
of three distinct regions: in the region to the left of the contact
discontinuity we have $\left(u_{\text{L}},\lambda_{\text{L}}\right)$,
between the contact and the shock the state is $\left(u_{\text{M}}=5,\lambda_{\text{R}}\right)$,
and to the right of the shock we have $\left(u_{\text{R}},\lambda_{\text{R}}\right)$.
As $u_{\text{R}}=D$, the characteristic equations in the rightmost
region are vertical. The shock speed in this case is $s=0.5$.

Figure \ref{fig:solutions-Riemann-problems}b shows the solution of
the Riemann problem for $u_{\text{L}}=4$, $u_{\text{R}}=6$, along
with the characteristic plane. In this case, the solution consists
of four distinct regions: to the left of the contact discontinuity,
$x=-Dt$, the solution is $\left(u_{\text{L}},\lambda_{\text{L}}\right)$,
between the contact and the tail of the rarefaction wave, the solution
is $\left(u_{\text{M}}=5,\lambda_{\text{R}}\right)$, inside the rarefaction
wave, we have $\left(u=x/t+D,\lambda_{\text{R}}\right)$, and to the
right of the head of the rarefaction solution, we have $\left(u_{\text{R}},\lambda_{\text{R}}\right)$.
The head and tail of the rarefaction wave are determined by the lines
$x=\left(u_{\text{R}}-D\right)t$ and $x=\left(u_{\text{M}}-D\right)t$,
respectively.

\subsection{Computation of detonation velocity\label{subsec:compute-D(t)}}

Now we require an algorithm to evolve $D$ in time. We use a combination
of the shock-evolution equation \eqref{eq:shock-evolution-equation}
for time steps when the flow is sufficiently smooth and the characteristics-based
method from \citep{kasimov2004dynamics} when the flow has steep gradients.
Note that in this subsection, subscript ``s'' is used for the staggered
grid point index $N+1/2$ for clarity.

At each time step, we estimate the velocity gradient $du/dx$ using
backward finite differences:
\begin{equation}
\left.\frac{du}{dx}\right|_{x_{i}}=\frac{u_{i}-u_{i-1}}{\DeltaX},
\end{equation}
and if $\|du/dx\|_{\infty}\leq10$, then the flow is considered smooth.
Otherwise, it is considered not smooth.

For smooth flow, we integrate the shock-evolution equation \eqref{eq:shock-evolution-equation}
in time simultaneously with (\ref{eq:model-Burgers}\textendash \ref{eq:model-reaction}),
approximating the velocity gradient appearing in \eqref{eq:shock-evolution-equation}
based on the following second-order approximation on the stencil $\{x_{N-1},x_{N},x_{\shk}\}$:
\begin{equation}
\left.\frac{du}{dx}\right|_{\shk}^{j}=\frac{u_{N-1}^{j}-9u_{N}^{j}+8u_{\shk}^{j}}{3\DeltaX}
\end{equation}
with $u_{\shk}^{j}=2D^{j}$ by the Rankine\textendash Hugoniot conditions.
Additionally, we set the CFL number to $0.4$.

For non-smooth flow, we use the method from \citep{kasimov2004dynamics}
slightly modifying it due to the use of a staggered grid in the present
computations. In this method, $D^{j+1}$ is computed using the forward-going
characteristic equation and then equations (\ref{eq:model-Burgers}\textendash \ref{eq:model-reaction})
are integrated in time; details of the algorithm are given below.
CFL number is set to 0.1 in this case.

The forward-going characteristic equation is
\begin{equation}
u\dot{u}+\sigma\dot{\lambda}-\sigma\omega(u,\lambda)=0\label{eq:Cplus-eq}
\end{equation}
satisfied along the trajectory
\begin{equation}
\dot{x}=u-D.\label{eq:Cplus-traj}
\end{equation}
At the time step $j+1$ we are given $u_{i}^{j}$, $\lambda_{i}^{j}$,
and $D^{j}$ for $i=0,\dots,N$ and aim to find $D^{j+1}$. We assume
that during time step $\DeltaT$ from $t^{j}$ to $t^{j+1}$, characteristics
are straight lines and then approximate the trajectory of the characteristic
that starts at location $x_{*}^{j}$ at time $t^{j}$ and arrives
to the shock $x_{\shk}=0$ at time $t^{j+1}$ using the forward Euler
method:
\begin{equation}
x_{\shk}-x_{*}^{j}=\left(u_{*}^{j}-D^{j}\right)\DeltaT,\label{eq:Cplu-traj-Euler}
\end{equation}
where $u_{*}^{j}=u(x_{*}^{j},t^{j})$ and $x_{\shk}=0$ for any time
step. To find $u_{*}^{j}$, we interpolate linearly using the value
at the shock and the given average of $u_{N}^{j}$ in the rightmost
computational cell:
\begin{equation}
\frac{u_{\shk}^{j}-u_{N}^{j}}{x_{\shk}-x_{N}}=\frac{u_{*}^{j}-u_{N}^{j}}{x_{*}-x_{N}}.
\end{equation}
Using relations $x_{\shk}-x_{N}=\DeltaX/2$ and $x_{N}=-\DeltaX/2$,
we arrive at the following formula for $u_{*}^{j}$:
\begin{equation}
u_{*}^{j}=u_{\shk}^{j}+\frac{2\left(u_{\shk}^{j}-u_{N}^{j}\right)}{\DeltaX}x_{*},
\end{equation}
with $u_{\shk}^{j}=2D^{j}$. Similarly, we obtain the formula for
$\lambda_{*}^{j}$:
\begin{equation}
\lambda_{*}^{j}=\lambda_{\shk}^{j}+\frac{2\left(\lambda_{\shk}^{j}-\lambda_{N}^{j}\right)}{\DeltaX}x_{*}
\end{equation}
with $\lambda_{\shk}^{j}=0$.

Now, by substituting expression for $u_{*}^{j}$ into \eqref{eq:Cplu-traj-Euler},
we find that $x_{*}^{j}$ can be computed explicitly:
\begin{equation}
x_{*}^{j}=\frac{D^{j}\Delta t}{-1-\frac{2\left(u_{\shk}^{j}-u_{N}^{j}\right)}{\DeltaX}}.
\end{equation}

After computing $x_{*}^{j}$, $u_{*}^{j}$, and $\lambda_{*}^{j}$,
the updated detonation velocity, $D^{j+1}$, is found by solving a
nonlinear equation arising from the discretization of \eqref{eq:Cplus-eq}:
\begin{equation}
u_{*}^{j}\frac{u_{\shk}^{j+1}-u_{*}^{j}}{\DeltaT}+\sigma\frac{\lambda_{\shk}^{j+1}-\lambda_{*}^{j}}{\DeltaT}-\sigma\omega\left(u_{\shk}^{j+1},\lambda_{\shk}^{j+1}\right)=0,\text{ where }u_{\shk}^{j+1}=2D^{j+1},\quad\lambda_{\shk}^{j+1}=0,
\end{equation}
which is done using the Newton method with the termination condition
that the relative error in $D^{j+1}$ between two subsequent iterations
is less than $10^{-8}$.

\subsection{Bifurcation diagram}

Now we turn our attention to various solutions of the model when the
activation energy $\theta$ is increased while other parameters are
kept fixed. For some values of $\theta$, the solution is found to
undergo bifurcations from one nonlinear regime to another. To visualize
these bifurcations, we run simulations for $\theta\in[0.90;1.15]$
with step $0.001$ up to time $\TFinal=1000$ and in each simulation
we extract local minima of the time series of detonation velocity,
$D(t)$, for time window $t\in[900;1000]$; subsequently, the minima
are plotted against $\theta$ on a bifurcation diagram. Each simulation
is conducted using $q=4$, $\tolLambda=10^{-6}$, and $\Nhrz=1280$
with an initial condition being the corresponding ZND solution. Then,
necessary perturbation of the initial condition is supplied by the
truncation error of the numerical scheme.

\begin{figure}
\begin{centering}
\includegraphics{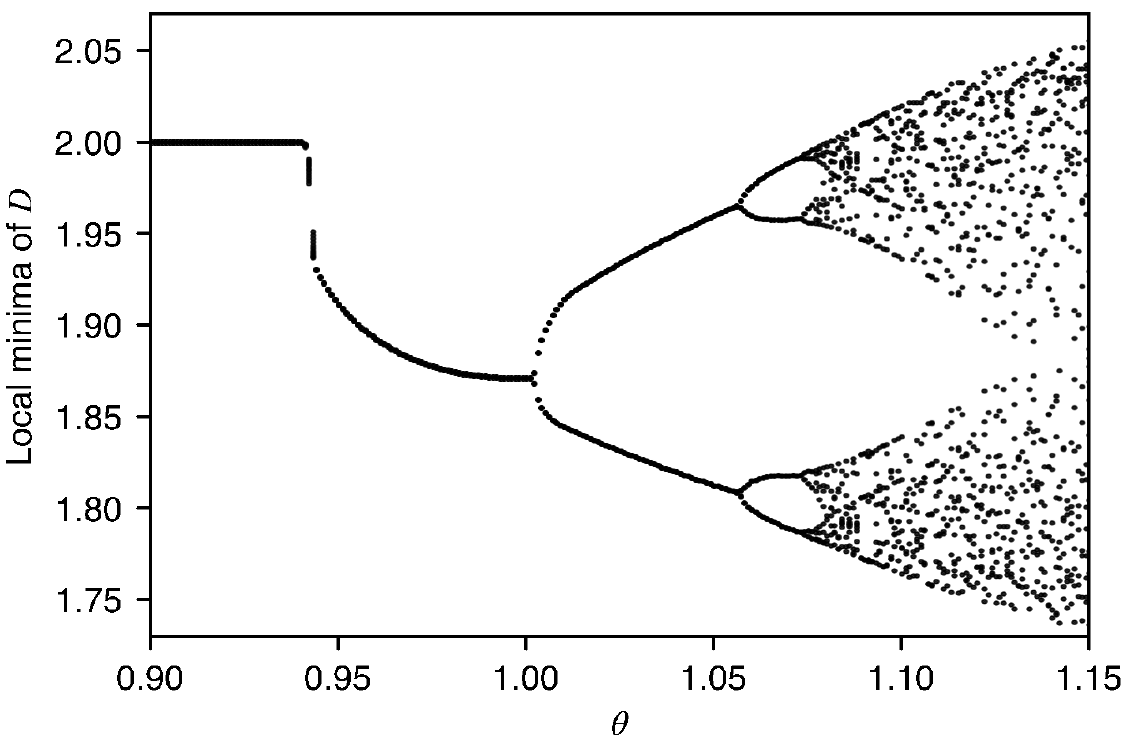}
\par\end{centering}
\caption{Bifurcation diagram for $q=4$, $\theta\in[0.95;1.15]$ with step
$\Delta\theta=0.001$.\label{fig:bif-diag}}
\end{figure}

Figure \ref{fig:bif-diag} shows the resultant bifurcation diagram.
Note its resemblance of the well-known diagram for the logistic map
in the sense that, as $\theta$ increases, solution undergoes a series
of period-doubling bifurcations until eventually chaotic-looking regimes
appear. Similar diagrams have also been found previously in various
models of detonation \citep{Ng2005,HenrickAslamPowers2006,romick2012effect,kasimovPRL2013,FariaKasimovRosales-JFM2015}. 

In Figure \ref{fig:bif-diag} at $\theta\in[0.9;0.936]$, the solution
is asymptotically stable as the initial perturbation decays with $D\to\DCJ$
as $t\to\infty$; at $\theta=0.937$ a bifurcation takes place from
a stable steady-state solution to a period-1 stable limit cycle, and
then the solution stays qualitatively the same for $\theta\in[0.937,1.001]$
with the local minima decreasing below $\DCJ$ value; at $\theta=1.002$
the second bifurcation occurs to a period-2 limit cycle, and then
the cycle is preserved for $\theta\in[1.002;1.056]$ with the top
minima increasing and bottom minima decreasing; at $\theta=1.074$
the solution bifurcates to a period-4 limit cycle. For larger $\theta$
($\theta\gtrapprox1.077$), the solution is appearing to be chaotic
as a large number of minima are found. However, for $\theta\in\{1.089,1.09\}$
period-6 limit cycle is obtained.

It must be mentioned that at $\theta\gtrapprox1.077$, it becomes
quite challenging to compute the solutions sufficiently accurately
due to internal shocks appearing and hitting the lead shock, effectively
making the reaction zone unresolved even for resolution $\Nhrz=1280$
that is used for these computations. We also run the simulations with
twice and four times coarser grids to verify the convergence of the
bifurcation diagram for stable limit cycles for $\theta\lessapprox1.077$.

\subsection{Stable limit cycles}

Now we consider time series for several different values of $\theta$
that correspond to qualitatively different nonlinear dynamics as evident
from the bifurcation diagram \ref{fig:bif-diag}. We also plot phase
portraits in $\dot{D}-D$ plane to better understand the dynamics.
The shock acceleration $\dot{D}$ is approximated by central finite
differences:
\begin{equation}
\dot{D}=\frac{D_{i+1}-D_{i-1}}{t_{i+1}-t_{i-1}}\text{ for }i=1,2,\dots,\label{eq:dD_dt_fd}
\end{equation}
and initial acceleration is taken to be zero, $\dot{D}\left(0\right)=0$.
As time series contain numerical noise, to regularize numerical differentiation,
the $D\left(t\right)$ series are first smoothed using the simple
moving average algorithm with window size 11, and after obtaining
$\dot{D}$ using \eqref{eq:dD_dt_fd}, it is also smoothed using the
same algorithm with window size 5.

\begin{figure}
\begin{centering}
\includegraphics{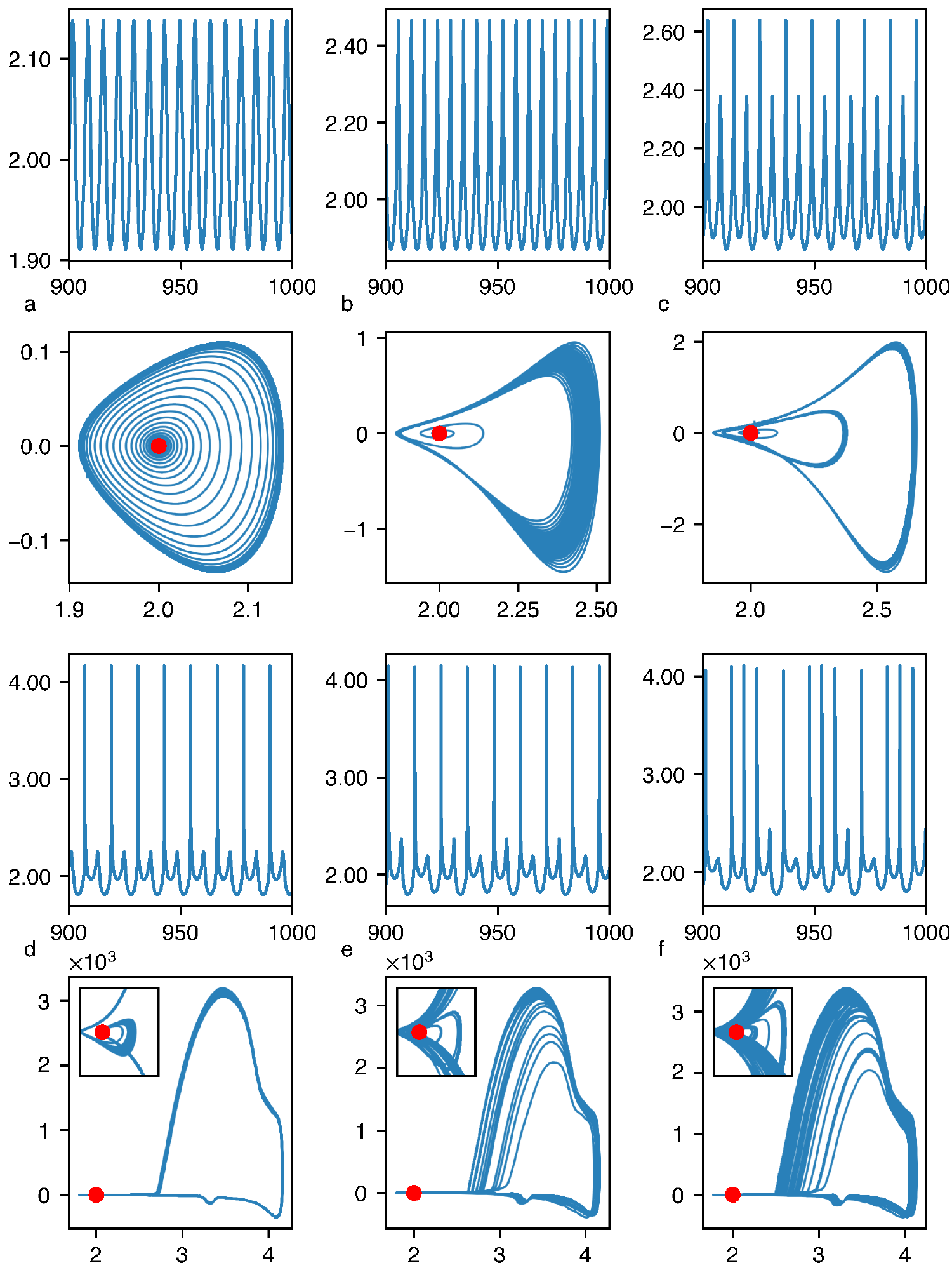}
\par\end{centering}
\caption{The time series and phase portraits of nonlinear solutions for $q=4$
and various $\theta$: (a) $\theta=0.95$, (b) $\theta=1$, (c) $\theta=1.004$,
(d) $\theta=1.055$, (e) $\theta=1.065$, (f) $\theta=1.089$. Plots
in the odd rows show time series $D$ versus $t$, while plots in
the even rows show corresponding phase portraits $\dot{D}$ versus
$D$. Insets in the bottom row have limits $(1.8,2.5)$ on the $x$-axis
and $(-1,1)$ on the $y$-axis.\label{fig:all-time-series}}
\end{figure}

Figures \ref{fig:all-time-series}a,b show the nonlinear dynamics
for $\theta=0.95$ and $\theta=1$, which is a stable limit cycle
in these two cases. Figures \ref{fig:all-time-series}c,d demonstrate
period-2 limit cycles for $\theta=1.004$ and $\theta=1.055$, respectively.
From these figures we can see that as $\theta$ increases, distance
between top and bottom minima increases as well: for $\theta=1.004$,
the ratio of relative minima is $\approx1.11$, while for $\theta=1.055$,
this ratio is $\approx1.84$. Figures \ref{fig:all-time-series}e,f
show period-4 nonlinear oscillations for $\theta=1.065$ and period-6
nonlinear oscillations for $\theta=1.089$. It can be also noticed
that as $\theta$ increases, the range of detonation acceleration
$\dot{D}$ increases dramatically: for weakly unstable cases $\dot{D}$
is in the order of unity, while for strongly unstable cases, it is
in the order of 1000 as can be seen from Figures \ref{fig:all-time-series}d-f.

\section{Code verification\label{sec:verif}}

In this section, we assess the correctness of our linear and nonlinear
solvers. It is widely accepted \citep{salari2000code,ROY2005} that
the most stringent test case is the comparison of the observable order
of accuracy with theoretical order of accuracy of the numerical methods
used to discretize the governing equations. For all convergence studies
below, we compute the observed order of accuracy by the following
procedure. Let the convergence study to be performed with $N$ grid
resolutions. Then the order of accuracy is 

\begin{equation}
r_{i}=\frac{\log\left(E_{i}/E_{i-1}\right)}{\log\left(\DeltaX_{i}/\DeltaX_{i-1}\right)},\quad i=3,\dots,N,\label{eq:order-of-accuracy}
\end{equation}
where $E_{i}$ is an error for the $i$-th resolution and $\DeltaX_{i}$
is a spatial step for the $i$-th grid resolution.

\subsection{Convergence study\label{subsub:verif:conv}}

We test the linear solver as follows. Simulations are run for $q=4$
and $\theta=1$ with several grid resolutions, and relative errors
are computed as
\begin{equation}
E_{i}=\frac{\|\psi'_{i}-\psi'_{i-1}\|}{\|\psi'_{i}\|},\quad i=2,\dots,N,\label{eq:fm:rel-errors}
\end{equation}
where $\psi'_{i}$ is the time series of the perturbation of detonation
velocity obtained with $i$-th grid resolution, $N$ is the total
number of resolutions, and norms are $L_{1}$, $L_{2}$, and $L_{\infty}$.
If the observed order of accuracy \eqref{eq:order-of-accuracy} matches
the theoretical fifth order, then one concludes that the implementation
of the solver is correct.

Table \ref{tab:verif:linear} shows the obtained errors and orders
of accuracy for the linear solvers where errors were computed with
different norms. It can be seen that the observed order of accuracy
is five, hence, the implementation of the linear solver is correct.
The sudden drop of the order of accuracy for the resolution $\Nhrz=1280$
is due to the domination of the rounding errors of the floating-point
arithmetic over the truncation errors of the numerical schemes at
this resolution.

\begin{table}
\centering{}\caption{Convergence study for the linear solver for $q=4$, $\theta=1$. $\protect\Nhrz$
is the resolution per half-reaction zone, $E_{1}$, $E_{2}$, $E_{\infty}$
are relative errors computed by \eqref{eq:fm:rel-errors} in $L_{1}$,
$L_{2}$, and $L_{\infty}$ norms, respectively; $r_{1}$, $r_{2}$,
$r_{\infty}$ are corresponding observed orders of accuracy. Negative
values for orders of accuracy are due to the domination of the round-off
error of floating-point arithmetic over truncation error of numerical
methods used.\label{tab:verif:linear}}
\begin{tabular*}{\textwidth}{@{\extracolsep{\fill}}ccccccc} \\
\toprule
$N_{1/2}$ & $E_1$ & $r_1$ & $E_2$ & $r_2$ & $E_\infty$ & $r_\infty$ \\
\midrule
20 & {N/A} & {N/A} & {N/A} & {N/A} & {N/A} & {N/A} \\
40 & \num{9e-05} & {N/A} & \num{9e-05} & {N/A} & \num{1e-04} & {N/A} \\
80 & \num{3e-06} & 5.20 & \num{3e-06} & 5.20 & \num{3e-06} & 5.19 \\
160 & \num{8e-08} & 5.10 & \num{8e-08} & 5.10 & \num{8e-08} & 5.10 \\
320 & \num{2e-09} & 5.06 & \num{2e-09} & 5.06 & \num{3e-09} & 5.06 \\
640 & \num{6e-11} & 5.42 & \num{7e-11} & 5.25 & \num{3e-10} & 3.13 \\
1280 & \num{6e-11} & -0.10 & \num{7e-11} & -0.05 & \num{4e-10} & -0.49 \\
\bottomrule
\end{tabular*}

\end{table}

For the nonlinear solver, we compute the solution on several different
grids for $q=4$, $\theta=0.92$. For these parameters, the solution
is stable, hence, $D\to\DCJ$ at large times. Here we take $\TFinal=300$.
We do not specify any perturbation in this case, so that the perturbation
of the initial traveling-wave solution is only due to the truncation
error of the scheme. The global error is then defined by
\[
E_{i}=\|D_{i}(t)-\DCJ\|_{2}\text{ for }i=1,\dots,N,
\]
where $D_{i}(t)$ is the time series of detonation velocity for the
$i$-th grid resolution, and the norm is $L_{2}$-norm. Then the observed
order of accuracy is computed using \eqref{eq:order-of-accuracy}.
Table \ref{tab:verif:nonlin-steady-state} shows the obtained errors
and the orders of accuracy. It can be seen that the order of accuracy
matches the theoretical second order which confirms that the nonlinear
solver described in Subsection \ref{subsec:riemann} is implemented
correctly and is second-order accurate at least for the stable flows.

\begin{table}
\caption{Convergence study for the nonlinear solver using parameters $q=4$,
$\theta=0.92$, $\protect\TFinal=300$.\label{tab:verif:nonlin-steady-state}}
\begin{tabular*}{\textwidth}{@{\extracolsep{\fill}}ccc} \\
\toprule
$N_{1/2}$ & $E$ & $r$ \\
\midrule
20 & \num{2e-04} & {N/A} \\
40 & \num{5e-05} & 2.11 \\
80 & \num{1e-05} & 2.05 \\
160 & \num{3e-06} & 2.01 \\
320 & \num{8e-07} & 1.99 \\
\bottomrule
\end{tabular*}

\end{table}

Additionally, we run a convergence study for the nonlinear solver
for the case when the solution is a stable limit cycle with small
amplitude of oscillations so that the solver uses only second-order
approximations. We run simulations for $q=4$ and $\theta=0.95$ with
several grid resolutions and extract local minima from the time series
of detonation velocity for $t\in[900;1000]$. For each grid resolution,
the average of the local minima $D_{\text{avg min},i}$ is computed.
Then the errors are computed by comparing the obtained average to
the next average:
\begin{equation}
E_{i}=|D_{\text{avg min},i}-D_{\text{avg min},i+1}|,\quad i=1,\dots N-1,
\end{equation}
where $N$ is the number of grid resolutions. Then the order of accuracy
is computed using \eqref{eq:order-of-accuracy}. Table \ref{tab:verif:nonlin-limit-cycle}
displays the obtained errors and orders of accuracy. It can be seen
that the observed order of accuracy is two as expected, hence the
implementation of the nonlinear solver is correct not only for stable
solutions, but also for weakly unstable solutions.

\begin{table}
\centering{}\caption{Convergence study for the nonlinear solver with parameters $q=4$,
$\theta=0.95$.\label{tab:verif:nonlin-limit-cycle}}
\begin{tabular*}{\textwidth}{@{\extracolsep{\fill}}ccc} \\
\toprule
$N_{1/2}$ & $E$ & $r$ \\
\midrule
20 & {N/A} & {N/A} \\
40 & \num{2e-03} & {N/A} \\
80 & \num{5e-04} & 2.13 \\
160 & \num{1e-04} & 2.07 \\
320 & \num{3e-05} & 2.04 \\
640 & \num{7e-06} & 2.03 \\
1280 & \num{2e-06} & 2.09 \\
\bottomrule
\end{tabular*}

\end{table}

\subsection{Comparison of linear and nonlinear solutions\label{subsec:comparison}}

In this subsection, we compare the solutions of the linear and nonlinear
problems for further verification. If parameters of the problem are
such that the ZND solution is unstable and the initial amplitude of
the perturbation of the solution is small relative to the ZND solution
itself, then for early times both linear and nonlinear solutions should
agree with each other.

\begin{figure}
\begin{centering}
\includegraphics{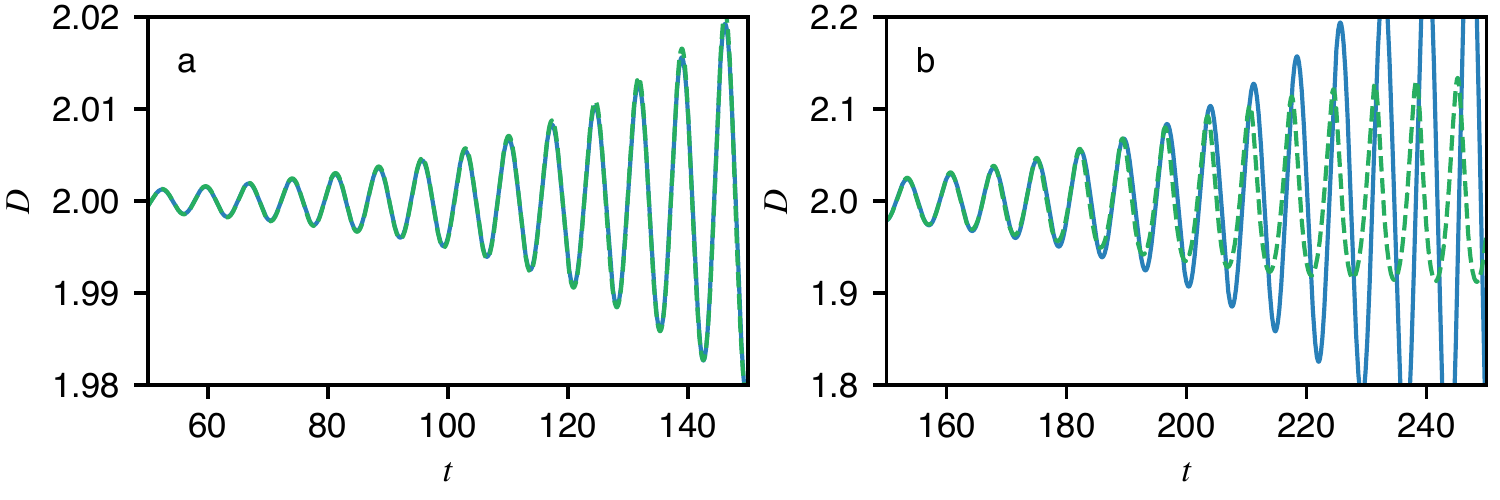}
\par\end{centering}
\caption{Comparison of solutions of linearized (solid line) and nonlinear (dashed
line) problems: a) at early times when the amplitude of perturbation
is small comparing to $D_{\mathrm{CJ}}=2$ and two solutions are indistinguishable;
b) at later times when the amplitude of perturbation is large and
two solutions diverge: linear solution continues to grow exponentially
while the nonlinear solution exhibits a limit cycle.\label{fig:comparison}}
\end{figure}

Figure \ref{fig:comparison} shows detonation velocity as a function
of time obtained both from linear and nonlinear solvers for heat release
$q=4$ and activation energy $\theta=0.95$, with resolutions $\Nhrz=40$
and $\Nhrz=320$ for linear and nonlinear solvers, respectively, and
initial perturbation amplitude $A_{0}=10^{-4}$. This choice of resolutions
and initial amplitude is dictated by the difference in the order of
accuracy of the solvers (fifth for linear and second for nonlinear).
In Figure \ref{fig:comparison}a, the detonation velocity is shown
at early times, when the amplitude of the perturbation is small. As
we can see, the solutions of the linear and nonlinear problems are
indistinguishable from each other: they both exhibit exponential growth
with the same frequency of oscillations. In Figure \ref{fig:comparison}b,
the detonation velocity is shown at later times when the amplitude
of the perturbation becomes comparable with $\DCJ$. It can be seen
that the linear solution continues to grow exponentially while the
nonlinear solution saturates on a limit cycle. The agreement of the
two solutions at early times additionally verifies the correctness
of both linear and nonlinear solvers.

\section{Conclusions\label{sec:conclusions}}

In this work, we carried out a comprehensive numerical investigation
of the linear and nonlinear dynamics of solutions of a relatively
simple $2\times2$ system of hyperbolic balance laws that possesses
nontrivial dynamical properties. It is demonstrated that traveling-wave
solutions of the system can become unstable as a system parameter
is varied. As a result of the instability, the solutions tend asymptotically
in time to a limit cycle attractor of varying complexity. Stable periodic
limit cycles as well as what appears to be a chaotic attractor are
found in the simulations. 

The numerical predictions are verified extensively by convergence
studies and comparisons of results obtained by different numerical
algorithms. The linear stability of traveling shock-wave solutions
is computed by a shock-fitting method which is fifth-order in space
and time using the direct method of \citep{kabanov2018linear}. These
results are compared with those computed by the method of normal modes,
and complete agreement is found. The nonlinear simulations are performed
using the second-order Godunov method implemented also in a shock-fitting
approach. Convergence tests and comparisons with linear predictions
when appropriate verify the accuracy of the computed results and confirm
the presence of limit cycles, period-doubling bifurcations, and possible
chaos in this simple hyperbolic system.

Extensions of the present work that are of theoretical interest include
problems involving additional factors, such as diffusive effects,
and understanding their role in the nature of the periodic and chaotic
solutions of the system. 

\section*{Supplementary Materials}

The datasets and the scripts reproducing the figures and tables in
this work are available at \href{https://doi.org/10.5281/zenodo.1297175}{https://doi.org/10.5281/zenodo.1297175}.

\section*{Acknowledgments}

DK is grateful to King Abdullah University of Science and Technology
(KAUST) for the financial support. AK was partially supported by the
Russian Foundation for Basic Research through grants \#17-53-12018
and \#17-01-00070. For computer time, this work used the resources
of the Supercomputing Laboratory at KAUST.

\bibliographystyle{elsarticle-harv}
\bibliography{fickett-model-stability.bbl}

\begin{thebibliography}{49}
\expandafter\ifx\csname natexlab\endcsname\relax\def\natexlab#1{#1}\fi
\expandafter\ifx\csname url\endcsname\relax
  \def\url#1{\texttt{#1}}\fi
\expandafter\ifx\csname urlprefix\endcsname\relax\def\urlprefix{URL }\fi

\bibitem[{Bdzil and Stewart(2007)}]{bdzil2007dynamics}
Bdzil, J.~B., Stewart, D.~S., 2007. The dynamics of detonation in explosive
  systems. Annu. Rev. Fluid Mech. 39, 263--292.

\bibitem[{Bdzil and Stewart(2012)}]{bdzil2012theory}
Bdzil, J.~B., Stewart, D.~S., 2012. Theory of detonation shock dynamics. In:
  Shock Waves Science and Technology Library, Vol. 6. Springer, pp. 373--453.

\bibitem[{Brown et~al.(1989)Brown, Byrne, and Hindmarsh}]{brown1989vode}
Brown, P.~N., Byrne, G.~D., Hindmarsh, A.~C., 1989. {VODE: A
  variable-coefficient ODE solver}. SIAM journal on scientific and statistical
  computing 10~(5), 1038--1051.

\bibitem[{Chen and Gurtin(1971)}]{ChenGurtin1971}
Chen, P., Gurtin, M., 1971. Growth and decay of one-dimensional shock waves in
  fluids with internal state variables. Physics of Fluids 14~(6), 1091--1094.

\bibitem[{Clavin(2017)}]{Clavin-CST-2017}
Clavin, P., 2017. Nonlinear dynamics of shock and detonation waves in gases.
  Combustion Science and Technology 189~(5), 747--775.
\newline\urlprefix\url{http://dx.doi.org/10.1080/00102202.2016.1260562}

\bibitem[{Clavin and Williams(2002)}]{clavin2002dynamics}
Clavin, P., Williams, F.~A., 2002. {Dynamics of planar gaseous detonations near
  Chapman-Jouguet conditions for small heat release}. Combustion Theory and
  Modelling 6~(1), 127--139.

\bibitem[{Clavin and Williams(2012)}]{clavin2012analytical}
Clavin, P., Williams, F.~A., 2012. Analytical studies of the dynamics of
  gaseous detonations. Philosophical Transactions of the Royal Society A:
  Mathematical, Physical and Engineering Sciences 370~(1960), 597--624.

\bibitem[{D{\"o}ring(1943)}]{Doering1943}
D{\"o}ring, W., 1943. Uber den detonationvorgang in gasen. Annalen der Physik
  43(6/7), 421--428.

\bibitem[{Faria and Kasimov(2015)}]{FariaKasimov-PROCI2014}
Faria, L.~M., Kasimov, A.~R., 2015. Qualitative modeling of the dynamics of
  detonations with losses. Proceedings of the Combustion Institute 35,
  2015--2023.
\newline\urlprefix\url{http://www.sciencedirect.com/science/article/pii/S1540748914003162}

\bibitem[{Faria et~al.(2014)Faria, Kasimov, and
  Rosales}]{FariaKasimovRosales-SIAM2014}
Faria, L.~M., Kasimov, A.~R., Rosales, R.~R., 2014. Study of a model equation
  in detonation theory. SIAM Journal on Applied Mathematics 74~(2), 547--570.
\newline\urlprefix\url{http://epubs.siam.org/doi/abs/10.1137/130938232}

\bibitem[{Faria et~al.(2015)Faria, Kasimov, and
  Rosales}]{FariaKasimovRosales-JFM2015}
Faria, L.~M., Kasimov, A.~R., Rosales, R.~R., 2015. {Theory of weakly nonlinear
  self-sustained detonations}. Journal of Fluid Mechanics 784, 163--198.

\bibitem[{Faria et~al.(2016)Faria, Kasimov, and
  Rosales}]{FariaKasimovRosales-SIAM2016}
Faria, L.~M., Kasimov, A.~R., Rosales, R.~R., 2016. Study of a model equation
  in detonation theory: multidimensional effects. SIAM J. Appl. Maths 76~(3),
  887--909.

\bibitem[{Fickett(1979)}]{Fickett1979}
Fickett, W., 1979. Detonation in miniature. American Journal of Physics
  47~(12), 1050--1059.
\newline\urlprefix\url{http://link.aip.org/link/?AJP/47/1050/1}

\bibitem[{Fickett(1984)}]{fickett1984shock}
Fickett, W., 1984. Shock initiation of detonation in a dilute explosive.
  Physics of Fluids 27, 94.

\bibitem[{Fickett(1985{\natexlab{a}})}]{Fickett1985}
Fickett, W., 1985{\natexlab{a}}. Introduction to Detonation Theory. University
  of California Press, Berkeley, CA.

\bibitem[{Fickett(1985{\natexlab{b}})}]{fickett1985stability}
Fickett, W., 1985{\natexlab{b}}. Stability of the square-wave detonation in a
  model system. Physica D: Nonlinear Phenomena 16~(3), 358--370.

\bibitem[{Fickett and Davis(2011)}]{FickettDavis2011}
Fickett, W., Davis, W.~C., 2011. Detonation: theory and experiment. Dover
  Publications.

\bibitem[{Gottlieb et~al.(2001)Gottlieb, Shu, and Tadmor}]{gottlieb2001strong}
Gottlieb, S., Shu, C., Tadmor, E., 2001. Strong stability-preserving high-order
  time discretization methods. SIAM review, 89--112.

\bibitem[{Hairer et~al.(1993)Hairer, N{\o}rsett, and
  Wanner}]{hairer1993solving}
Hairer, E., N{\o}rsett, S.~P., Wanner, G., 1993. {Solving Ordinary Differential
  Equations I: Nonstiff problems}. Springer.

\bibitem[{Henrick et~al.(2006)Henrick, Aslam, and
  Powers}]{HenrickAslamPowers2006}
Henrick, A.~K., Aslam, T.~D., Powers, J.~M., 2006. Simulations of pulsating
  one-dimensional detonations with true fifth order accuracy. J. Comput. Phys.
  213~(1), 311--329.

\bibitem[{Humpherys et~al.(2013)Humpherys, Lyng, and
  Zumbrun}]{humpherys2013stability}
Humpherys, J., Lyng, G., Zumbrun, K., 2013. {Stability of viscous detonations
  for Majda's model}. Physica D: Nonlinear Phenomena 259, 63--80.

\bibitem[{Humpherys and Zumbrun(2010)}]{humpherys2010efficient}
Humpherys, J., Zumbrun, K., 2010. Efficient numerical stability analysis of
  detonation waves in {ZND}. arXiv preprint arXiv:1011.0897.

\bibitem[{Jones et~al.(2001--)Jones, Oliphant, Peterson, et~al.}]{scipy}
Jones, E., Oliphant, T., Peterson, P., et~al., 2001--. {SciPy}: Open source
  scientific tools for {Python}.
\newline\urlprefix\url{http://www.scipy.org/}

\bibitem[{Kabanov and Kasimov(2018)}]{kabanov2018linear}
Kabanov, D.~I., Kasimov, A.~R., 2018. Linear stability analysis of detonations
  via numerical computation and dynamic mode decomposition. Physics of Fluids
  30~(3), 036103.

\bibitem[{Kasimov et~al.(2013)Kasimov, Faria, and Rosales}]{kasimovPRL2013}
Kasimov, A.~R., Faria, L.~M., Rosales, R.~R., 2013. Model for shock wave chaos.
  Physical Review Letters 110~(10), 104104.

\bibitem[{Kasimov and Stewart(2004)}]{kasimov2004dynamics}
Kasimov, A.~R., Stewart, D.~S., 2004. On the dynamics of self-sustained
  one-dimensional detonations: A numerical study in the shock-attached frame.
  Physics of Fluids 16, 3566.

\bibitem[{Lee and Stewart(1990)}]{LeeStewart90}
Lee, H.~I., Stewart, D.~S., 1990. Calculation of linear detonation instability:
  One-dimensional instability of plane detonation. J. Fluid Mech. 212,
  103--132.

\bibitem[{LeVeque(1992)}]{leveque1992numerical}
LeVeque, R., 1992. Numerical methods for conservation laws. Birkh{\"a}user.

\bibitem[{LeVeque(2002)}]{leveque2002finite}
LeVeque, R.~J., 2002. Finite volume methods for hyperbolic problems. Cambridge
  University Press.

\bibitem[{Levy(1992)}]{levy1992majda}
Levy, A., 1992. On {Majda}'s model for dynamic combustion. Communications in
  partial differential equations 17~(3-4), 657--698.

\bibitem[{Lyng and Zumbrun(2004)}]{lyng2004stability}
Lyng, G., Zumbrun, K., 2004. A stability index for detonation waves in
  {M}ajda's model for reacting flow. Physica D: Nonlinear Phenomena 194~(1),
  1--29.

\bibitem[{Majda(1980)}]{Majda1980}
Majda, A., 1980. A qualitative model for dynamic combustion. SIAM Journal on
  Applied Mathematics 41~(1), 70--93.
\newline\urlprefix\url{http://link.aip.org/link/?SMM/41/70/1}

\bibitem[{Ng et~al.(2005)Ng, Higgins, Kiyanda, Radulescu, Lee, Bates, and
  Nikiforakis}]{Ng2005}
Ng, H., Higgins, A., Kiyanda, C., Radulescu, M., Lee, J., Bates, K.,
  Nikiforakis, N., 2005. Nonlinear dynamics and chaos analysis of
  one-dimensional pulsating detonations. Combust. Theory Model 9~(1), 159--170.

\bibitem[{Radulescu and Tang(2011)}]{RadulescuPRL2011}
Radulescu, M.~I., Tang, J., 2011. Nonlinear dynamics of self-sustained
  supersonic reaction waves: Fickett's detonation analogue. Phys. Rev. Lett.
  107~(16).

\bibitem[{Romick et~al.(2012)Romick, Aslam, and Powers}]{romick2012effect}
Romick, C.~M., Aslam, T.~D., Powers, J.~M., 2012. The effect of diffusion on
  the dynamics of unsteady detonations. Journal of Fluid Mechanics 699, 453.

\bibitem[{Rosales(1989)}]{rosales1989diffraction}
Rosales, R.~R., 1989. Diffraction effects in weakly nonlinear detonation waves.
  In: Nonlinear Hyperbolic Problems. Vol. 1402 of Lecture Notes in Mathematics.
  Springer, pp. 227--239.

\bibitem[{Rosales and Majda(1983)}]{RosalesMajda1983}
Rosales, R.~R., Majda, A.~J., 1983. Weakly nonlinear detonation waves. SIAM
  Journal on Applied Mathematics 43~(5), 1086--1118.

\bibitem[{Roy(2005)}]{ROY2005}
Roy, C.~J., 2005. Review of code and solution verification procedures for
  computational simulation. Journal of Computational Physics 205~(1), 131 --
  156.
\newline\urlprefix\url{http://www.sciencedirect.com/science/article/pii/S0021999104004619}

\bibitem[{Salari and Knupp(2000)}]{salari2000code}
Salari, K., Knupp, P., 2000. Code verification by the method of manufactured
  solutions. Tech. rep., Sandia National Labs., Albuquerque, NM (US); Sandia
  National Labs., Livermore, CA (US).

\bibitem[{Schmid(2010)}]{schmid2010dynamic}
Schmid, P.~J., 2010. Dynamic mode decomposition of numerical and experimental
  data. Journal of Fluid Mechanics 656, 5--28.

\bibitem[{Stewart and Kasimov(2005)}]{StewartKasimovSIAP05}
Stewart, D.~S., Kasimov, A.~R., 2005. Theory of detonation with an embedded
  sonic locus. SIAM J. Appl. Maths. 66~(2), 384--407.

\bibitem[{Stewart and Kasimov(2006)}]{stewart2006state}
Stewart, D.~S., Kasimov, A.~R., 2006. State of detonation stability theory and
  its application to propulsion. Journal of Propulsion and Power 22~(6), 1230.

\bibitem[{Tang and Radulescu(2012)}]{tang2012dynamics}
Tang, J., Radulescu, M., 2012. Dynamics of shock induced ignition in {F}icketts
  model: Influence of $\chi$. Proceedings of the Combustion Institute.

\bibitem[{Taylor et~al.(2009)Taylor, Kasimov, and Stewart}]{taylor2009mode}
Taylor, B.~D., Kasimov, A.~R., Stewart, D.~S., 2009. Mode selection in weakly
  unstable two-dimensional detonations. Combustion Theory and Modelling 13~(6),
  973--992.

\bibitem[{Trefethen and Bau~III(1997)}]{trefethen1997numerical}
Trefethen, L.~N., Bau~III, D., 1997. Numerical linear algebra. Vol.~50. Siam.

\bibitem[{Tu et~al.(2014)Tu, Rowley, Luchtenburg, Brunton, and
  Kutz}]{tu2014dynamic}
Tu, J.~H., Rowley, C.~W., Luchtenburg, D.~M., Brunton, S.~L., Kutz, J.~N.,
  2014. {On dynamic mode decomposition: theory and applications}. Journal of
  Computational Dynamics 1~(2), 391--421.

\bibitem[{von Neumann(1942)}]{vonNeumann1942}
von Neumann, J., 1942. Theory of detonation waves. {O}ffice of {S}cientific
  {R}esearch and {D}evelopment, {R}eport 549. Tech. rep., National Defense
  Research Committee Div. B.

\bibitem[{Zel'dovich(1940)}]{Zeldovich1940}
Zel'dovich, Y.~B., 1940. On the theory of propagation of detonation in gaseous
  systems. J. Exp. Theor. Phys. 10~(5), 542--569.

\bibitem[{Zumbrun(2017)}]{zumbrun2017recent}
Zumbrun, K., 2017. Recent results on stability of planar detonations. In:
  Shocks, Singularities and Oscillations in Nonlinear Optics and Fluid
  Mechanics. Springer, pp. 273--308.

\end{thebibliography}

\end{document}